\documentclass[twocolumn]{aastex701}

\newcommand{\anst}{\mathring{\mathrm{A}}}
\newcommand{\HeII}{He\,{\sc ii}}

\begin{document}

\title{Little Red Dots as Supermassive Counterparts of SS~433: A Hyper-Eddington Accretion Framework}
\shorttitle{LRDs as Supermassive SS~433 Counterparts}

\author[orcid=0009-0005-2801-6594]{Shuying Zhou}
\affiliation{Department of Astronomy, Xiamen University, Xiamen, Fujian 361005, People’s Republic of China; msun88@xmu.edu.cn}
\email{zhoushuying@stu.xmu.edu.cn}

\author[orcid=0000-0002-0771-2153]{Mouyuan Sun} 
\affiliation{Department of Astronomy, Xiamen University, Xiamen, Fujian 361005, People’s Republic of China; msun88@xmu.edu.cn}
\email[show]{msun88@xmu.edu.cn}  

\author[0000-0002-1660-9502]{Xihan Ji}
\affiliation{Kavli Institute for Cosmology, University of Cambridge, Madingley Road, Cambridge CB3 0HA, UK}
\affiliation{Cavendish Laboratory, University of Cambridge, 19 JJ Thomson Avenue, Cambridge CB3 0HE, UK}
\email{xj274@cam.ac.uk}

\author[0000-0002-7329-9344]{Ya-Ping Li}
\affiliation{Shanghai Astronomical Observatory, Chinese Academy of Sciences, 80 Nandan Road, Shanghai 200030, People’s Republic of China}
\email{liyp@shao.ac.cn}

\author[0000-0001-6947-5846]{Luis C. Ho}
\affiliation{Kavli Institute for Astronomy and Astrophysics, Peking University, Beijing 100871, People’s Republic of China}
\affiliation{Department of Astronomy, School of Physics, Peking University, Beijing 100871, People’s Republic of China}
\email{lho.pku@gmail.com}

\author[orcid=0000-0002-4985-3819]{Roberto Maiolino}
\affiliation{Kavli Institute for Cosmology, University of Cambridge, Madingley Road, Cambridge CB3 0HA, UK}
\affiliation{Cavendish Laboratory, University of Cambridge, 19 JJ Thomson Avenue, Cambridge CB3 0HE, UK}
\affiliation{Department of Physics and Astronomy, University College London, Gower Street, London WC1E 6BT, UK}
\email{RM665@cam.ac.uk}

\author[0000-0002-4223-2198]{Zhen-Yi Cai}
\affiliation{Department of Astronomy, University of Science and Technology of China, Hefei, Anhui 230026, People’s Republic of China}
\affiliation{School of Astronomy and Space Science, University of Science and Technology of China, Hefei, Anhui 230026, People’s Republic of China}
\email{zcai@ustc.edu.cn}

\author[0000-0002-1530-2680]{Hai-Cheng Feng}
\affiliation{Yunnan Observatories, Chinese Academy of Sciences, Kunming 650216, Yunnan, People's Republic of China}
\affiliation{Key Laboratory for the Structure and Evolution of Celestial Objects, Chinese Academy of Sciences, Kunming 650216, Yunnan, People's Republic of China}
\affiliation{Center for Astronomical Mega-Science, Chinese Academy of Sciences, 20A Datun Road, Chaoyang District, Beijing 100012, People's Republic of China}
\affiliation{Key Laboratory of Radio Astronomy and Technology, Chinese Academy of Sciences, 20A Datun Road, Chaoyang District, Beijing 100101, People's Republic of China}
\email{hcfeng@ynao.ac.cn}

\author[orcid=0009-0008-7583-5658]{Manqi Fu}
\affiliation{Department of Astronomy, Xiamen University, Xiamen, Fujian 361005, People’s Republic of China; msun88@xmu.edu.cn}
\email{fumanqi@stu.xmu.edu.cn}

\author[orcid=0000-0003-3137-1851]{Wei-Min Gu} 
\affiliation{Department of Astronomy, Xiamen University, Xiamen, Fujian 361005, People’s Republic of China; msun88@xmu.edu.cn}
\email{guwm@xmu.edu.cn}  

\author[orcid=0000-0001-8678-6291]{Tong Liu} 
\affiliation{Department of Astronomy, Xiamen University, Xiamen, Fujian 361005, People’s Republic of China; msun88@xmu.edu.cn}
\email{tongliu@xmu.edu.cn}

\author[orcid=0000-0003-4874-0369]{Junfeng Wang} 
\affiliation{Department of Astronomy, Xiamen University, Xiamen, Fujian 361005, People’s Republic of China; msun88@xmu.edu.cn}
\email{jfwang@xmu.edu.cn}

\author[orcid=0000-0001-7349-4695]{Jianfeng Wu} 
\affiliation{Department of Astronomy, Xiamen University, Xiamen, Fujian 361005, People’s Republic of China; msun88@xmu.edu.cn}
\email{wujianfeng@xmu.edu.cn}

\author[0000-0002-1935-8104]{Yongquan Xue}
\affiliation{Department of Astronomy, University of Science and Technology of China, Hefei, Anhui 230026, People’s Republic of China}
\affiliation{School of Astronomy and Space Science, University of Science and Technology of China, Hefei, Anhui 230026, People’s Republic of China}
\email{xuey@ustc.edu.cn}

\begin{abstract}

High-redshift little red dots (LRDs) are compact sources characterized by V-shaped spectral energy distributions (SEDs), broad emission lines, and often prominent Balmer breaks. Their high number density and apparently large black hole masses suggest that they are essential to the early evolution of galaxies and supermassive black holes (SMBHs); however, the nature of their central engines remains uncertain. Here, we propose that LRDs are the supermassive, high-redshift counterparts of the hyper-Eddington accreting Galactic microquasar SS~433, viewed at high inclinations. By scaling the hyper-Eddington accretion physics from stellar-mass black holes to supermassive scales, we show that the observed LRD features, including X-ray weakness, soft optical SEDs, apparent sub-Eddington accretion ratio, and Balmer breaks, emerge naturally from the self-shielding geometry of a puffed-up accretion disk. In this framework, the broad-line regions are ionized by anisotropic radiation escaping from the inner disk, analogous to the unseen UV/X-ray emission revealed by the W50 nebula in SS 433. Their low-inclination or lower-accretion-rate counterparts would appear as little blue dots (LBDs) or normal active galactic nuclei. Our model predicts that the Balmer break strength positively correlates with the broad-line width, that the emission lines are more variable than the optical continuum, that LRDs are intrinsically more luminous than observed, and that LBDs are more variable than LRDs. This unified-scale model redefines LRDs as the essential laboratories for observing the rapid accretion-driven growth that shaped the early assembly of galaxies and their central SMBHs. 

\end{abstract}


\section{Introduction}\label{sect:intro}
The James Webb Space Telescope (JWST) has revealed numerous high-redshift galaxies, revising our view of galaxy formation and supermassive black hole (SMBH) evolution. Among them, mysterious little red dots (LRDs) are compact sources with red optical continua, V-shaped spectral energy distributions (SEDs), broad emission lines (BELs), and weak X-ray emission \citep[e.g.,][]{Harikane2023, Kocevski2023, Perna2023, Greene2024,deGraaff2025, Naidu2025, JiXH2025a, ZhuangMY2026}. Some of them exhibit an extremely strong Balmer break that cannot be explained by typical stellar populations \citep[e.g.,][]{de_Graaff2025-cliff, JiXH2025a, Naidu2025, Ivey2026}. Several local LRD analogs have also been identified \citep[e.g.,][]{JiXH2026, Euclid2026, LinXJ2026, Lin2026_llrd}. Given their apparently large single-epoch black hole masses ($M_{\mathrm{BH}}$) and high number density, LRDs may provide important clues to early SMBH seeding and growth, and their coevolution with host galaxies \citep[e.g.,][]{Harikane2023, Bogdan2024, Maiolino2024, Akins2025, LiJY2025}. 

LRDs have attracted extensive theoretical attention \citep[for a recent summary, see, e.g.,][]{IL2025}, and most models invoke SMBH accretion (see Table~\ref{tab:model}). Mildly super-Eddington accretion may explain the X-ray weakness \citep[e.g.,][]{Pacucci2024, Madau2024}. The initial explanation for the red optical continuum is the heavy dust extinction; moreover, the lack of submillimetre detection \citep{Akins2025, Chen2025} indicates that the dust is compact, possibly circumnuclear \citep[e.g.,][]{Madau2026, Pacucci2026}. Alongside dust reddening, the red optical continuum can also be attributed to either photospheric emission from spherical inflow \citep[e.g.,][]{Begelman2008, LiuHP2025, Begelman2026}, dense broad-line region (BLR) gas \citep[e.g.,][]{JiXH2025a, Madau2026}, or stellar processes in self-gravity-dominated disk regions \citep[e.g.,][]{Zhang2026, Wang2025b, Chen2026}. Furthermore, the origin of the UV emission remains unclear, with potential contributors from the accretion disk itself \citep[e.g.,][]{Labbe2024, Zhang2026, Wang2025b, Ando2026, Ji2026_lya}, scattered disk light, the host galaxy \citep[e.g.,][]{Greene2024,WangBJ2024,Naidu2025}, and the nebula emission \citep[e.g.,][]{CHChen2025}. The prominent Balmer break has been attributed to dense gas in the form of a spherical envelope \citep{Begelman2008, Kido2025, LiuHP2025, Begelman2026} or BLR clouds along the line of sight \citep[e.g.,][]{Inayoshi2025, JiXH2025a, Maiolino2025}. While the super-Eddington accretion is often favored, the physical nature of LRDs remains unsettled. 

SS~433 is probably the only known persistent super-Eddington accretor \citep[e.g.,][]{Fabrika2007}. According to recent mass determinations, the compact object in SS~433 is a stellar-mass black hole with $M_{\mathrm{BH,433}}\simeq 5$--$9\ M_{\odot}$ \citep{Cherepashchuk2019}. Its dimensionless accretion rate\footnote{We define $\dot{m}\equiv \dot{M}/\dot{M}_{\mathrm{Edd}}$, with $\dot{M}_{\mathrm{Edd}}=10L_{\mathrm{Edd}}/c^2$, and $L_{\mathrm{Edd}}$ is calculated by assuming the opacity of $\kappa=0.34\ \mathrm{cm^2 g^{-1}}$.} is $\dot{m}\gtrsim 10^{-4}\ M_{\odot}\ \mathrm{yr}^{-1}/(2.2\times 10^{-8} (M_\mathrm{BH}/M_{\odot})\ M_{\odot}\ \mathrm{yr^{-1}})\simeq 500$, placing SS~433 in the hyper-Eddington regime \citep[e.g.,][]{Cherepashchuk1981}. The inclination angle of SS~433 with respect to the orbital axis is $i\simeq 78^{\circ}$ \citep[i.e., nearly edge on; see,][]{Margon1989, Eikenberry2001}. Interestingly, the disk UV/optical emission ($3000\ \text{\AA}$--$7000\ \text{\AA}$) of SS~433 can be well modeled by a blackbody function with the effective temperature of $5\times 10^{4}$--$7\times 10^{4}$ K \citep{Dolan1997}, much softer than the theoretical expectation ($\sim 10^7$ K). In addition, SS~433 is undetectable at UV wavelengths of $\lesssim2000\ \text{\AA}$ \citep{Gies2002}. \textit{The physical mechanism that drives the soft SED in SS~433 may also be responsible for the red optical continuum observed in LRDs.}

\begin{figure*}
\centering
\includegraphics[width=1.\linewidth]{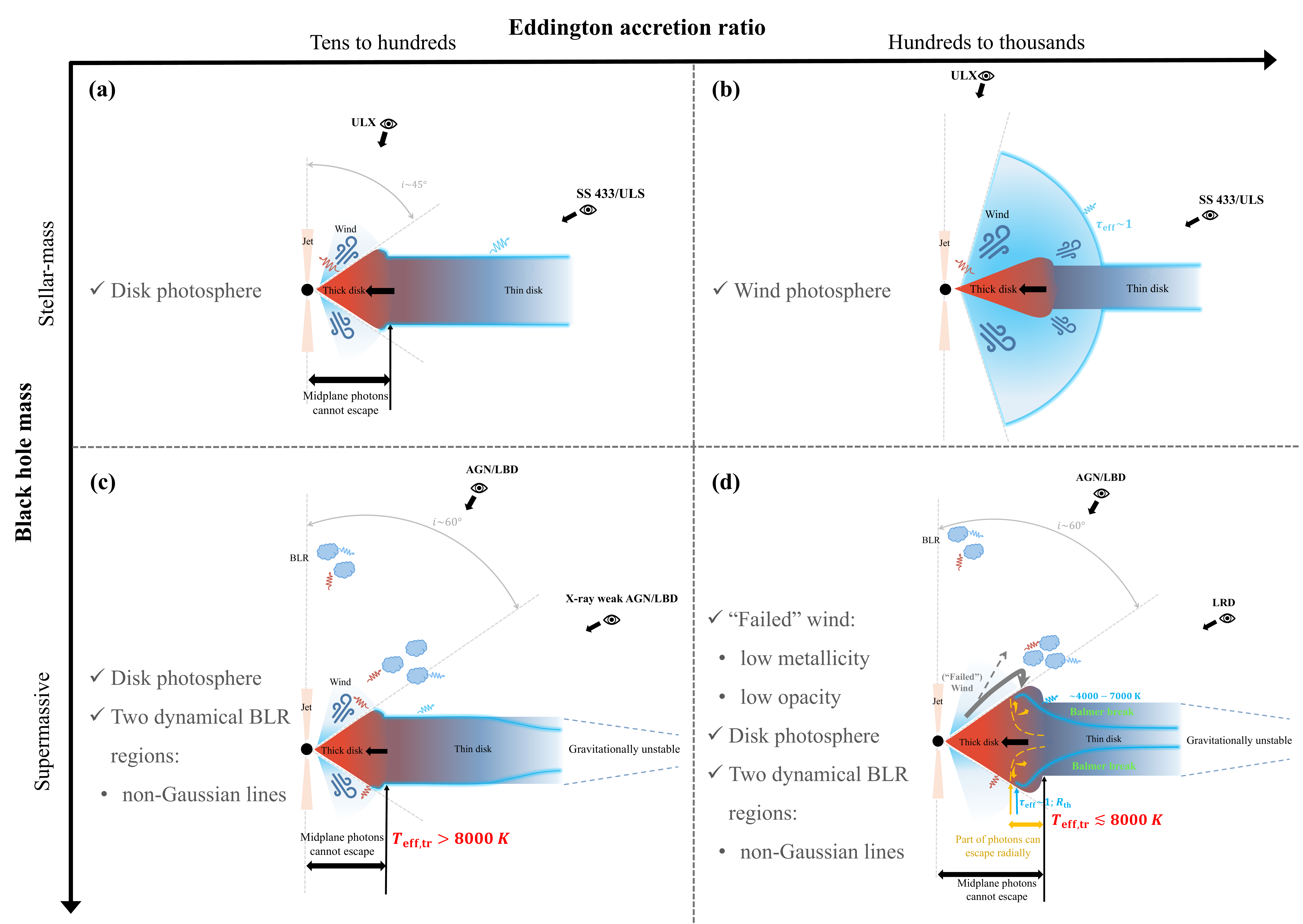}
\caption{Schematic view of the central-engine models for LRDs, SS~433, and ULSs across different black hole masses and Eddington accretion ratios. For stellar-mass black holes, a hyper-Eddington accretion disk with $T_{\mathrm{eff}}\gg 10^{4}$ K and opacity $\sim 0.34\ \mathrm{cm^2\ g^{-1}}$ can launch strong radiation-driven winds. At $\dot{m}\gtrsim 500$ (panel b), the wind is optically thick and yields a soft SED, as in SS~433 and some ULSs; at $\dot{m}\simeq 10$ (panel a), the wind is optically thin, but self-shielding still allows either hard or soft SEDs depending on inclination. Similar behavior is expected for SMBHs with $M_{\mathrm{BH}}=10^6$--$10^8\ M_{\odot}$ at $\dot{m}\simeq 10$ (panel c). At $\dot{m}\gtrsim 500$ (panel d), winds would only be present in the hot inner regions due to the radial evolution of the local Eddington limit. In the cool outer disk ($T \lesssim 8000$ K), sharp decreases in opacity increase the local Eddington limit; therefore, strong winds are suppressed. The low opacity also results in the outer disk being partially transparent, producing the optical continuum and Balmer break. Conversely, in the hot inner regions ($T \gtrsim 10^4$ K), the reduced local Eddington limit caused by increased opacity inevitably triggers powerful radiation-pressure-driven winds. As these inner winds expand and cool below the recombination temperature, the radiation pressure weakens, causing a substantial fraction of the material to fall back onto the disk (i.e., ``failed winds'') and rejoin the accretion flow. At high inclinations, the puffed-up inner disk geometrically self-shields the hot inner emission, resulting in an LRD SED. At low inclinations, the full thick-disk emission is visible, yielding an LBD-like SED. The BLR clouds are photoionized by the highly anisotropic inner ionization field.} \label{fig1-model}
\end{figure*}

Previous observational \citep[e.g.,][]{Heuvel1981} and theoretical \citep[e.g.,][]{King2000} studies have suggested that SS~433 launches strong winds with a mass outflow rate comparable to the accretion rate. Such winds can be optically thick, with a large scattering optical depth $\tau_{\mathrm{es}}\sim 100$, and thus soften the underlying disk emission throughout the wind \citep[panel b in Figure~\ref{fig1-model}; e.g.,][]{Begelman2006}. The wind photosphere then sets the observed disk SED, which is approximately a blackbody with $T_{\mathrm{eff}}\sim 10^5$ K. This SED is much softer than expected from a bare accretion disk. The same picture has also been invoked for ultraluminous X-ray sources \citep[ULXs, seen close to face-on; for a recent review, see, e.g.,][]{Kaaret2017, King2023} and their supersoft counterparts at high inclinations, i.e., ultraluminous supersoft sources \citep[ULSs; e.g.,][]{King2003, Poutanen2007, Shen2015, Soria2016}. 

The accretion rate of a black hole can far exceed the Eddington accretion limit $\dot{M}_{\mathrm{Edd}}$ (defined by assuming an radiative efficiency of $0.1$ and the opacity of $\kappa=0.34\ \mathrm{cm^2 g^{-1}}$), and advection due to ``photon trapping'' \citep{Begelman1978} is important \cite[see, e.g.,][]{Abramowicz1988, Zhang2025}. The flow should therefore consist of an outer standard thin disk \citep[SSD;][]{SSD} and a geometrically thick inner disk (hereafter the SLIM disk) inside a truncation radius $R_{\mathrm{tr}}$. This radius, which increases with $\dot{m}$ \citep{Abramowicz1988}, is often close to the ``spherization'' radius defined by \cite{SSD}, where strong winds are launched. For self-consistency, the thermalization radius in the wind scenario must exceed $R_{\mathrm{tr}}$, implying an outflow rate of $\dot{m}\gtrsim 500$ \citep[e.g.,][]{Soria2016}. If $\dot{m}\ll 500$, the winds should be too optically thin to thermalize the high-energy disk emission. Even then, self-shielding by the puffed-up thick disk can block the hard photons, so a ULX may still appear as a ULS at $i\gtrsim 45^{\circ}$ \citep[panel a in Figure~\ref{fig1-model};][]{Gu2016}. SS~433 also fits into this picture because of its high inclination and hyper-Eddington accretion rate. In reality, strong winds and the advection in SLIM disks should jointly determine the observational appearance of SS~433 \citep[e.g.,][]{Poutanen2007}. 

Observations of SS 433 and its surrounding W50 nebula provide convincing evidence that its hyper-Eddington accretion produces highly anisotropic UV and X-ray emission. First, optical spectra of gas filaments in the W50 nebula exhibit \HeII, H$\beta$, and [O\,{\sc iii}] emission lines, indicating strong unseen ionizing continua of $\sim 10^{40}\ \mathrm{erg\ s^{-1}}$ from SS~433 \citep{Fabrika2008-w50}, which is about two orders of magnitude larger than the detected optical emission from the edge-on sightline. Second, broadband X-ray observations reveal the strong unseen X-ray emission from SS~433 \citep{Middleton2021}. Hence, while an observer at a high inclination only sees the self-shielded soft red continuum, gas in polar regions is fully exposed to high-energy ionizing radiation escaping from the innermost regions. The same anisotropy may also be found in hyper-accretion SMBHs. 

Motivated by the softer-than-expected emission and extreme anisotropic radiation in SS~433, we propose a hyper-Eddington accretion structure for SMBHs in LRDs by scaling the underlying accretion physics from the stellar-mass to the supermassive regime. The soft emission in LRDs manifests as the observed red optical continuum, which can be roughly fitted by a modified blackbody with $T_{\mathrm{eff}}\approx5000$--$7000$ K \citep[see e.g.,][for potential revisions of the blackbody model]{deGraaff2025}. Meanwhile, the anisotropic ionizing radiation, analogous to the unseen UV/X-ray emission in SS~433, illuminates the BLR clouds, naturally producing the BELs in LRDs. The radiative transfer of dense but cold gas \citep[e.g.,][]{LiuHP2025} at the disk photosphere can create the strong non-stellar Balmer break observed in some LRDs \citep[e.g.,][]{JiXH2025a}. We therefore interpret LRDs as hyper-Eddington SMBHs viewed at high inclination angles.

The manuscript is organized as follows: Section~\ref{sect:idea} describes our hyper-Eddington accretion model, its comparisons with observations, and implications; Section~\ref{sect:dis} compares our model with others and explores its broader physical implications. The main conclusions are summarized in Section~\ref{sect:sum}. We adopt the flat $\Lambda \mathrm{CDM}$ cosmology with $H_0=70\ \mathrm{km\ s^{-1}\ Mpc^{-1}}$, $\Omega_\mathrm{m}=0.3$, and $\Omega_\Lambda=0.7$ in the manuscript. We denote $r=R/R_{\mathrm{S}}$ as the radial distance from the SMBH in units of the Schwarzschild radius $R_{\mathrm{S}}=2GM_{\mathrm{BH}}/c^2$. 

\section{The idea}\label{sect:idea}
\subsection{Hyper-Eddington accretion: from stellar-mass to supermassive black holes}
\label{sect:framework}
Following the framework in SS~433, we consider a hyper-Eddington accretion flow consisting of an outer standard SSD and an inner SLIM disk within $R_{\mathrm{tr}}$ (panel d in Figure~\ref{fig1-model}). Due to the significant advection cooling and intense radiation pressure at hyper-Eddington accretion rates, the inner SLIM disk is puffed up to be geometrically thick, with a scale height $H$ comparable to the radius $R=rR_{\mathrm{S}}$. Analytic work suggests that $H$ can even exceed $R$ \citep[e.g.,][]{Gu2009,Gu2012}. We define $R_{\mathrm{tr}}$ as the photon-trapping radius, which is determined by the condition that the vertical diffusion timescale, $t_{\mathrm{diff}} = 3H\tau_{\mathrm{mid}}/c$, is equal to the accretion timescale, $t_{\mathrm{acc}} = R/V_R$, where $\tau_{\mathrm{mid}}$, $V_R$ and $c$ denote the midplane optical depth, the radial velocity and speed of light, respectively. In the electron scattering dominated regime ($\kappa_{\mathrm{tot}}\simeq \kappa_{\mathrm{es}}$), the transition radius scales as $R_{\mathrm{tr}}=15\dot{m}(H/R)R_{\mathrm{S}}\simeq 15\dot{m}R_{\mathrm{S}}$ \citep[][]{Wang1999, Kato2008}. For the outer SSD region, the effective temperature profile is given by
\begin{equation}
    \label{eq:T_SSD}
    T_{\mathrm{eff,out}} =  \left( \frac{3GM_{\mathrm{BH}}\dot{M}f}{8\pi \sigma_{\mathrm{SB}}R^3} \right)^{1/4},
\end{equation}
we set the factor $f=1 - \sqrt{R_{\mathrm{ISCO}}/R}\simeq 1$ in the following analysis because $R\gg R_{\mathrm{ISCO}}$ (with $R_{\mathrm{ISCO}}=3R_{\mathrm{S}}$ for a Schwarzschild black hole). At the truncation radius $R=R_{\mathrm{tr}}$, the effective temperature is 
\begin{equation}\label{eq:Ttr}
T_{\mathrm{eff,tr}}=8000\ [\mathrm{K}] \times (M_{\mathrm{BH}}/[10^6 M_{\odot}])^{-0.25}(\dot{m}/1000)^{-0.5} \\.    
\end{equation}
If mass loss is negligible, the effective temperature profile of the inner SLIM disk is flatter than that of the SSD because a significant fraction of the viscous heat is advected into the black hole rather than being radiated. The effective temperature of the inner SLIM disk can be expressed as \citep[e.g.,][]{Kato2008}
\begin{equation}
    \label{eq:T_SLIM}
    T_{\mathrm{eff,in}} = T_{\mathrm{eff,tr}} \left(\frac{R}{R_{\mathrm{tr}}}\right)^{-0.5} = 10^6\ [\mathrm{K}]\ r^{-0.5} \left(\frac{M_{\mathrm{BH}}}{10^6 M_{\odot}}\right)^{-0.25}.
\end{equation}
Note that the accretion rate of the SLIM disk is assumed to be constant. In reality, the inner SLIM disk can drive powerful jets, and a considerable fraction of the accretion power may be released as the mechanical energy of the jets. Consequently, the effective temperature should be lower than that given by Eq.~\ref{eq:T_SLIM}. Hence, comparing with a hyper-accretion stellar black hole with $M_{\mathrm{BH}} \sim 10 M_{\odot}$, the same disk with $M_{\mathrm{BH}} \gtrsim 10^6 M_{\odot}$ has a much lower temperature and can barely produce sufficient hard ionizing photons ($\gtrsim 50\ \mathrm{eV}$) to excite strong high-ionization lines such as \HeII\ \citep[e.g.,][]{Tang2025, Brazzini2026}, albeit it can generate $\sim 10\ \mathrm{eV}$ to ionized hydrogen. Meanwhile, the self-shielding structure effectively reduces the ionizing continuum that illuminates the high-inclination BLR clouds, further eliminating the high-ionization BELs. 

Previous studies suggest that SMBH accretion at $\dot{m}\sim 500$ can overcome radiative feedback \citep[e.g.,][]{Inayoshi2016} and may not drive strong winds if the infalling gas has sufficiently low specific angular momentum \citep{Begelman2017}. Consequently, the disk remains geometrically thick out to thousands of $R_\mathrm{S}$, where the surface temperature at truncation radius $R_\mathrm{tr}$ drops to $T_{\mathrm{eff,tr}}=8000\ [\mathrm{K}]$ (Eq.~\ref{eq:Ttr}), triggering hydrogen recombination. This recombination causes both electron-scattering and true-absorption opacities to drop sharply \citep[$\kappa\sim T^{\sim 13}$; see, e.g.,][]{LiuHP2025}. Because the Eddington luminosity is inversely proportional to the opacity ($L_{\mathrm{Edd}} \propto \kappa^{-1}$), this sharp decline in opacity (by up to four orders of magnitude) results in great increases of the local Eddington limit in the cool outer disk. The radiation pressure in the disk surface of this outer region is therefore insufficient to launch powerful winds, allowing the outer disk to sustain a massive inflow rate without significant mass loss (also see the cold regions in the simulations of \citet{Toyouchi2024}. Conversely, as the gas accretes inward and the temperature rises above the hydrogen recombination threshold ($T \gtrsim 10^4$ K), the opacity rises back toward the electron-scattering-dominated value ($\kappa \sim 0.34\ \mathrm{cm^2\ g^{-1}}$), lowering the local Eddington limit. The inner regions therefore may launch strong radiation-pressure-driven winds, as originally proposed by \cite{SSD} \citep[also see in hydrodynamical simulations, e.g.,][]{Jiang2019, Toyouchi2024, Zhang2025}. However, as these inner winds expand and cool to the hydrogen recombination temperature, they would experience strong deceleration due to the rapid decrease of radiation pressure. A fraction of the decelerated winds may form BLR clouds, while the others fall back and rejoin the accretion flow. In such a system, while the mass accretion rate in the outer disk is large, its value in the SMBH event horizon may be greatly reduced to be around $\dot{M}_{\mathrm{Edd}}$ \citep[e.g.,][]{Toyouchi2024}. 

Furthermore, while line-driven winds could potentially exist, their strength depends critically upon the gas metallicity \citep{Gayley1995}. Thus, in low-metallicity environments, which are potentially associated with high-redshift LRDs \citep[e.g.,][]{Trefoloni2025, Ivey2026, Maiolino2026}, the line force multiplier is significantly reduced, resulting in a much lower mass-loss rate than in high-metallicity environments \citep{Castor1975, Proga1999}. 

The expectation of wind absence is consistent with the lack of significant, large-scale winds observed in high-redshift AGNs \citep{Maiolino2025}. In addition, the BLR reverberation mapping of super-Eddington-accreting black holes in the local universe reveals a lack of universal blueshifted wind signatures, with some instead exhibiting evidence of infalling gas in the BLR \citep[e.g.,][]{Du2016}, which also supports the idea that winds are not universally produced in SMBH super-Eddington systems. Consequently, the wind-photosphere model invoked for SS~433\footnote{The self-shield disk may also contribute to the observed soft spectrum.} and ULSs cannot be directly extrapolated to LRDs. Instead, we unify these systems within a broader hyper-Eddington framework in which the UV/optical continuum of LRDs is governed by geometric self-shielding from the inflated SLIM disk itself, rather than by an optically thick wind.

At high inclination angles, the puffed-up geometry of the inner SLIM disk self-shields the innermost accretion flow significantly (panel d of Figure~\ref{fig1-model}). This leads to highly anisotropic and inclination-dependent SEDs, as discussed for BHXRBs \citep{Watarai2005} and AGNs \citep{Wang2014}. At large inclination angles, the disk SED is dominated by the outer regions. We use the wind framework (see their section~5.2) of \cite{Soria2016} and the \texttt{NuPac} opacity code (also see Section~\ref{sect:cont}) of \cite{Morag2026} to determine the photoshphere properties; we find that the thermalized winds are expected to have blackbody temperatures of $10^4$--$10^5$ K and photoshphere radii of only $\sim 10^3\ R_\mathrm{S}$ (i.e., 
close to $0.1 R_{\mathrm{tr}}$) for SMBHs with $M_\mathrm{BH}\sim10^7M_\odot$ and $\dot{m}\sim 1000$. Thus, while the inner winds may be present, they cannot contribute to the high-inclination SEDs in our model. The same self-shielding effect has also been proposed to explain the observed X-ray weakness of LRDs \citep[e.g.,][]{Madau2024}, although those models are based on the much thicker non-advective ``Polish doughnut'' \citep[e.g.,][]{PB1980} rather than the SLIM disk (for more discussion of the differences between our model and the ``Polish doughnut'' model, see Section~\ref{sect:comp}). Meanwhile, inner winds, probably present, can also absorb the X-ray photons. In such models, the innermost X-ray and extreme-ultraviolet emission are obscured unless the line of sight is nearly face-on. Furthermore, the intrinsic X-ray emission is expected to be suppressed at high accretion rates due to enhanced Compton cooling \citep[e.g.,][]{Done2007, Madau2024}. 

According to Eq.~\ref{eq:Ttr}, we expect a typical LRD temperature of $5\times 10^4\ {(M_{\mathrm{BH,LRD}}/M_{\mathrm{BH,433}})}^{-0.25}C\ [\mathrm{K}]\simeq 3000\,{(M_{\mathrm{BH,LRD}}/[10^6 M_{\odot}])}^{-0.25}C\ [\mathrm{K}]$, where $C={(r_{\mathrm{tr,LRD}}/r_{\mathrm{tr,433}})}^{-0.5}$. For LRDs hosting SMBHs with $M_{\mathrm{BH}}\approx O(10^6) M_{\odot}$ \cite[e.g.,][]{Wang2025b}, our disk model naturally accounts for the $\sim 5000$ K optical continua (for details, see Section~\ref{sect:cont}). With such a cold temperature, the dense gaseous disk is responsible for the Balmer break (for details, see Section~\ref{sect:balmer}). 

\subsection{The red optical continuum}\label{sect:cont}
To evaluate the maximum observable temperature of the disk, we use a 1D toy model focusing on the advective photon trapping radius ($R_\mathrm{adv}$), where the radial photon diffusion timescale is equal to the accretion timescale ($t_{\mathrm{diff, radial}}\simeq t_{\mathrm{acc}}$). For an extremely large inclination angle (i.e., $\sim90^\circ$), the surface emission from inner regions is advected to the central SMBH, resulting in $R_\mathrm{adv}$ being the innermost visible radius. At that radius, the corresponding surface temperature is the maximum observable temperature. The radial diffusion timescale is expressed as $t_{\mathrm{diff, radial}}\simeq 3R(\kappa_{\mathrm{R}}\rho R)/c$ \citep{Kato2008}, where $\kappa_{\mathrm{R}}$ and $\rho=\dot{M}/(4\pi R H |V_\mathrm{R}|)$ denote the total Rosseland-mean opacity (i.e., the gray approximation) of the SLIM disk and the gas density at radius $R$, respectively. We use the \texttt{NuPac} code \citep{nupac, Morag2026} to generate a high-frequency-resolution and Rosseland-mean opacity table with a low metallicity of $Z=0.005Z_\odot$ \citep{Trefoloni2025, Ivey2026, LinXJ2026, Maiolino2026}. The opacity table covers temperatures from $58$ K to $9.3\times10^7$ K and densities from $10^{-22}$ to $\mathrm{10^{-4}\ g\ cm^{-3}}$. We use bilinear interpolation in logarithmic space to obtain the opacity at arbitrary density-temperature pairs. The accretion timescale is estimated by $t_{\mathrm{acc}}\simeq R/|V_\mathrm{R}|$, where $|V_\mathrm{R}|\simeq \beta \sqrt{GM_{\mathrm{BH}}/R}$ is the radial inflow velocity. The value of $\beta$ is uncertain and depends on the viscous transport efficiency within the accretion disk; here, we adopt $\beta=0.044$ \citep{Wang1999}. Hence, $\kappa_{\mathrm{R}}=(c/(3\beta \sqrt{GM_{\mathrm{BH}}/R}))(1/(\rho R))$. 

\begin{figure}
\includegraphics[width=1.\linewidth]{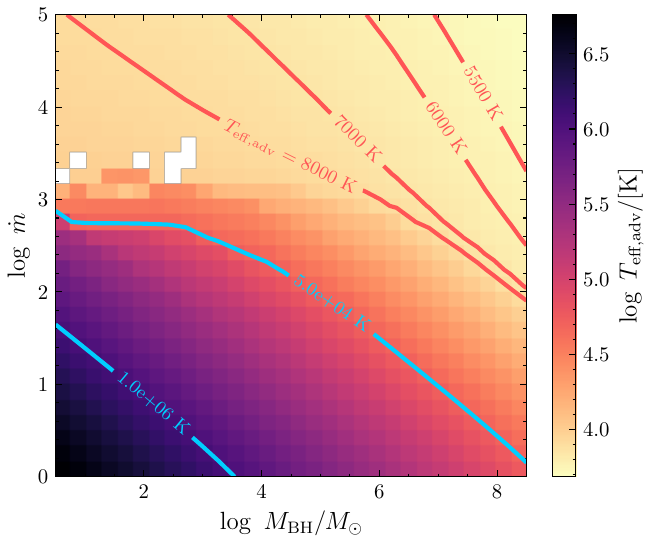}
\caption{Disk surface temperature at the innermost radius ($R_\mathrm{adv}$) visible along a high-inclination sightline (i.e., $\sim90^\circ$). At this radius, the advection timescale equals the radial photon diffusion timescale. Note that the solutions in the white pixels are numerically unstable. The two cyan curves correspond to the observed effective temperatures of SS~433 \citep{Fabrika2007} and ULSs. The four red curves show typical effective temperatures for LRDs. For SMBHs with $\dot{m}\gtrsim 500$, the maximum visible inner-disk temperature is $5000$--$8000$ K, similar to that inferred for LRDs. \label{fig2-teff}} 
\end{figure}

For a given $M_{\mathrm{BH}}$ and $\dot{m}$, we obtain the characteristic temperature by solving $t_{\mathrm{diff, radial}}=t_{\mathrm{acc}}$ (Figure~\ref{fig2-teff}). In the stellar-mass black hole regime, we obtain $10^4$ ($10^6$) K for $\dot{m}\sim 500$ ($\dot{m}\sim 30$), in agreement with SS~433 and ULSs, respectively. We note that for stellar-mass black holes, the dramatic change of opacity due to the partially ionized hydrogen may potentially cause the solution to be unstable for $\dot{m}\sim 1500$--$4000$. In the SMBH regime, the opacity $\kappa_{\mathrm{R}}$ becomes extremely sensitive to the gas temperature \citep[with $\kappa \sim T^{\sim 13}$; see, e.g.,][]{LiuHP2025}. Consequently, the maximum observable effective temperature at high inclination depends only weakly on $\rho$ (and thus on $\dot{m}$) or $M_{\mathrm{BH}}$. For SMBHs with $\dot{m}\gtrsim 500$, this temperature is constrained to $5000$--$8000$~K, consistent with LRD observations \citep[see, e.g., figure~3 in][]{deGraaff2025}. The geometric self-shielding effectively obscures the UV emission from the inner disk with $T > 8000$~K, and the sharp drop in opacity restricts the maximum observable temperature for SMBHs to $5000$--$8000$~K, thereby naturally producing the red optical continuum in LRDs. 

If the inclination angle is low and/or $\dot{m}\ll 100$ for SMBHs, one can directly detect emission from inner regions. Then, the system appears as little blue dots \citep[LBDs;][]{Brazzini2026}, which have a blue SED, high-ionization BELs, and lack the Balmer break (see Section~\ref{sect:balmer}). LBDs should be X-ray weak due to high Eddington accretion ratios. 

\begin{figure*}
\centering
\includegraphics[width=.9\linewidth]{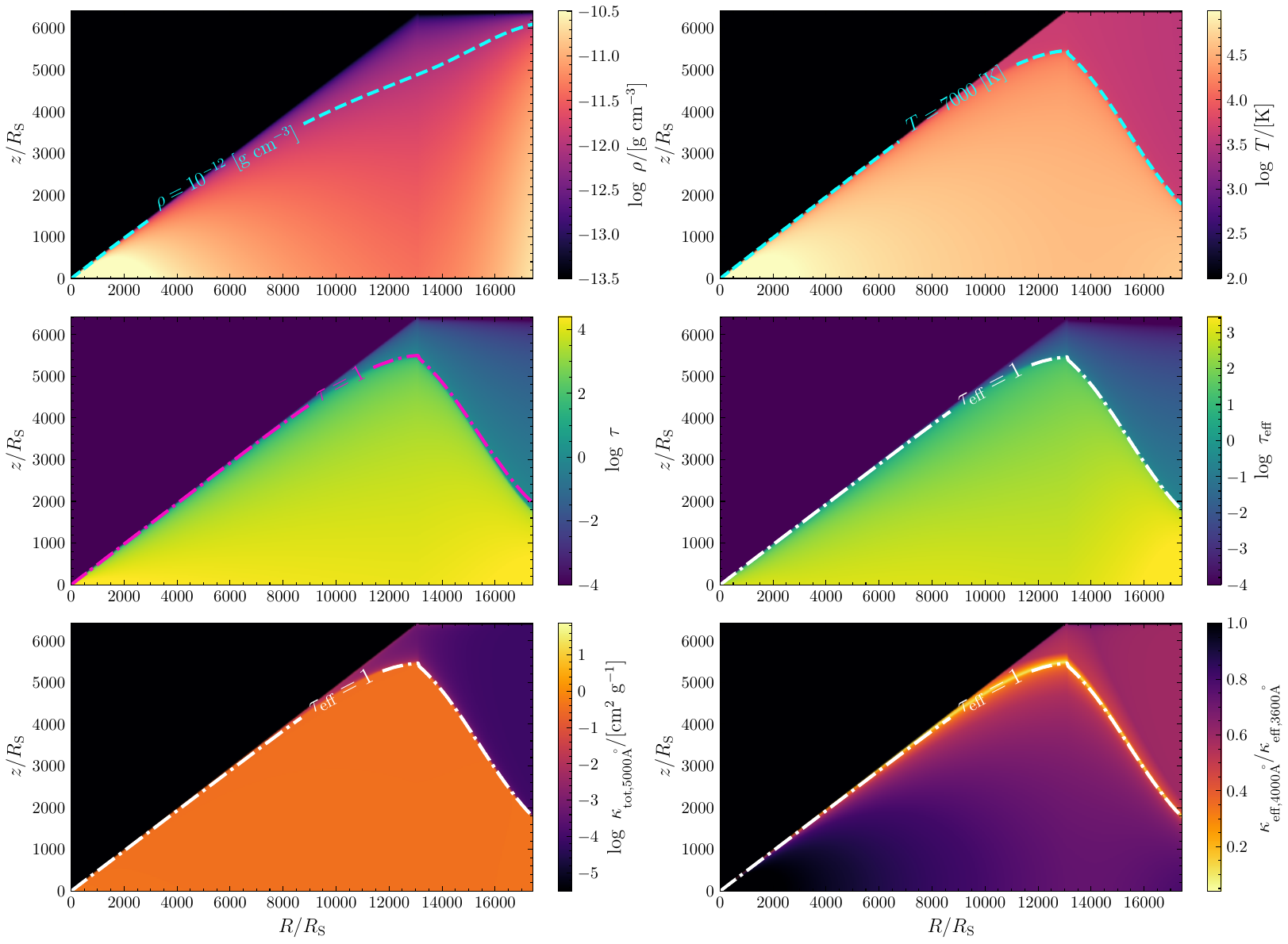}
\caption{Vertical structure of a hyper-Eddington accretion disk around an SMBH with $M_{\mathrm{BH}}= 10^7 M_{\odot}$ and $\dot{m}=1000$. The six panels show the gas density ($\rho$), temperature ($T$), face-on total optical depth ($\tau$), face-on effective optical depth ($\tau_\mathrm{eff}$), total opacity at $5000\ \text{\AA}$ ($\kappa_\mathrm{tot, 5000\text{\AA}}$), and the effective opacity ratio between $4000\ \text{\AA}$ and $3600\ \text{\AA}$ ($\kappa_\mathrm{eff, 4000\text{\AA}}/\kappa_\mathrm{eff, 3600\text{\AA}}$), respectively. The ratio $\kappa_\mathrm{eff, 4000\text{\AA}}/\kappa_\mathrm{eff, 3600\text{\AA}}$ serves as a proxy for the production of Balmer breaks; the region of maximum Balmer break production (indicated by the yellow color in the final panel) aligns closely with the thermalization surface ($\tau_\mathrm{eff}=1$). \label{fig3-vs}}
\end{figure*}

\subsection{Emergent SED and the Balmer break}\label{sect:balmer}
As pointed out by a number of previous works \citep[e.g.,][]{Inayoshi2025, JiXH2025a, LiuHP2025, Begelman2026}, the Balmer break observed in some LRDs can be naturally attributed to the rapid decline in opacity between the rest-frame wavelengths of $3600\ \anst$ and $4000\ \anst$ for a temperature of $\lesssim 10^4$ K. This is because hydrogen atoms with a principal quantum number of $n=2$ efficiently absorb photons with rest-frame wavelengths shorter than the Balmer limit of $\lambda\approx3646\ \anst$. Quantifying this effect requires solving the radiative transfer equations within the hyper-Eddington accretion disk. 

As a first step, we obtain the vertical structure of the accretion disk by solving the hydrostatic equilibrium, energy balance, and radiative transfer equations (see Appendix~\ref{appendix:vs}). In our calculations, we adopt a fiducial value of the dimensionless viscosity $\alpha=0.1$ unless otherwise specified. An example of the resulting vertical structure is shown in Figure~\ref{fig3-vs}. Here, the total and effective optical depths are defined as $\tau_{\mathrm{tot}}=\int_{z}^{H}\kappa_{\mathrm{R}}\rho dz$ and $\tau_{\mathrm{eff}}=\int_{z}^{H}\sqrt{\kappa_{\mathrm{R}}\kappa_{\mathrm{abs}}}\rho dz$ with $\kappa_{\mathrm{abs}}$ is the true absorption opacity, respectively. A substantial fraction of the outer disk is optically thin, with $\tau_{\mathrm{tot}}<1$ or $\tau_{\mathrm{eff}}<1$, and is therefore also transparent to emission from adjacent inner regions. The thermalization surface ($\tau_{\mathrm{eff}}=1$) is located in a regime with gas densities of $\sim 10^{-12}$--$10^{-11}\ \mathrm{g\ cm^{-3}}$ and temperatures of $\sim7000$ K. As shown in the final panel of Figure~\ref{fig3-vs}, this thermalization surface closely aligns with the region where the opacity ratio $\kappa_\mathrm{eff, 4000\text{\AA}}/\kappa_\mathrm{eff, 3600\text{\AA}}$ reaches its minimum. As this ratio serves as a proxy for Balmer break production, its alignment with the thermalization surface ensures that the emergent SED is shaped by the sharp opacity discontinuity at the Balmer limit, providing a robust mechanism for producing prominent Balmer breaks.

Building upon the derived disk vertical structures, we compute the emergent SED by solving the frequency-dependent radiative transfer equation in a two-dimensional $(r,z)$ configuration. We assume local thermodynamic equilibrium (LTE) and consider true absorption and isotropic electron scattering. The radiative transfer equation is given by
\begin{equation}
    \frac{dI_\nu}{ds}=\chi_\nu(S_\nu-I_\nu),
\end{equation}
where $I_\nu$ is the specific intensity, $S_\nu$ is the source function, $ds$ is the path length along a ray, and $\chi_\nu=\rho(\kappa_\mathrm{abs,\nu}+\kappa_\mathrm{es})$ is the total extinction coefficient. Under LTE and in the presence of scattering, the source function is coupled to the mean intensity $J_\nu$ via $S_{\nu}=(1-\epsilon_{\nu})J_{\nu}+\epsilon_{\nu}B_{\nu}$, where $\epsilon_{\nu}=\kappa_{\mathrm{abs,\nu}}/(\kappa_{\mathrm{abs,\nu}}+\kappa_{\mathrm{es}})$ and $B_{\nu}$ is the Planck function. Detailed descriptions are provided in Appendix~\ref{appendix:RT}.

\begin{figure}
\centering
\includegraphics[width=1.\linewidth]{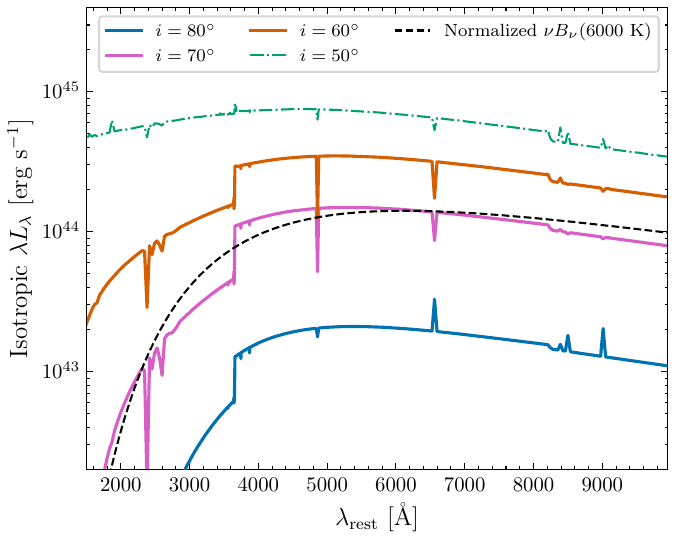}
\caption{Emergent SED for an SMBH with $M_{\mathrm{BH}}= 10^7 M_{\odot}$ and $\dot{m}=1000$ at different inclinations, obtained through radiative transfer simulations with the disk vertical structure in Figure~\ref{fig3-vs}. While the spectra are well-approximated by a normalized $6000\ \mathrm{K}$ blackbody (black dashed curve), a distinct Balmer break emerges at high inclinations ($\gtrsim 60^{\circ}$). \label{fig4-spec}}
\end{figure}

The observed monochromatic luminosity for a given inclination $i$ is obtained by integrating the emergent intensity over the visible surfaces of the near and far (relative to the observer) sides of the disk (see Appendix~\ref{appendix:RT}). The resulting SED for an SMBH with $M_{\mathrm{BH}}= 10^7 M_{\odot}$ and $\dot{m}=1000$ is shown in Figure~\ref{fig4-spec}. The spectrum is well-approximated by a blackbody with $T_{\mathrm{eff}}= 6000$ K, and the Balmer break is clearly visible in high-inclination cases with $i\gtrsim 60^{\circ}$. This is due to severe geometric obscuration of the inner disk at large angles. The high-temperature UV emission is blocked by the ``puffed-up'' disk, making the Balmer break from the cooler outer radii more visible. 

Our radiative transfer models also produce emission lines (e.g., SED with an inclination angle of $80^\circ$ in Figure~\ref{fig4-spec}). Within the scattering-dominated layers, the underlying continuum radiation is significantly diluted; however, the emission lines remain optically thick and effectively thermalize with the local gas. Consequently, the emission lines stand out against the diluted local continuum. A similar process was also proposed for cataclysmic variables \citep[e.g.,][]{Williams1980}. We expect the emission lines originating from the accretion disk to have broad profiles, with velocity widths of several thousand kilometers per second caused by Keplerian rotation, and slight shifts induced by gas inflow from near (redshifted) and far (blueshifted) sides (see Appendix~\ref {appendix:RT}). However, we note that the exact line fluxes and profiles of these lines depend upon non-LTE effects, which are beyond the scope of this work. We expect that these lines are weaker than BELs from BLRs and are undetectable. 

While our static radiative transfer calculations yield relatively sharp spectral features at the Balmer limit, kinematic broadening from gas turbulence in the accretion disk can smooth these edges. Accurately reproducing the intricate features of observed spectra requires the incorporation of kinematic information into radiative transfer frameworks, which is beyond the scope of this work.

\begin{figure*}
\centering
\includegraphics[width=.8\linewidth]{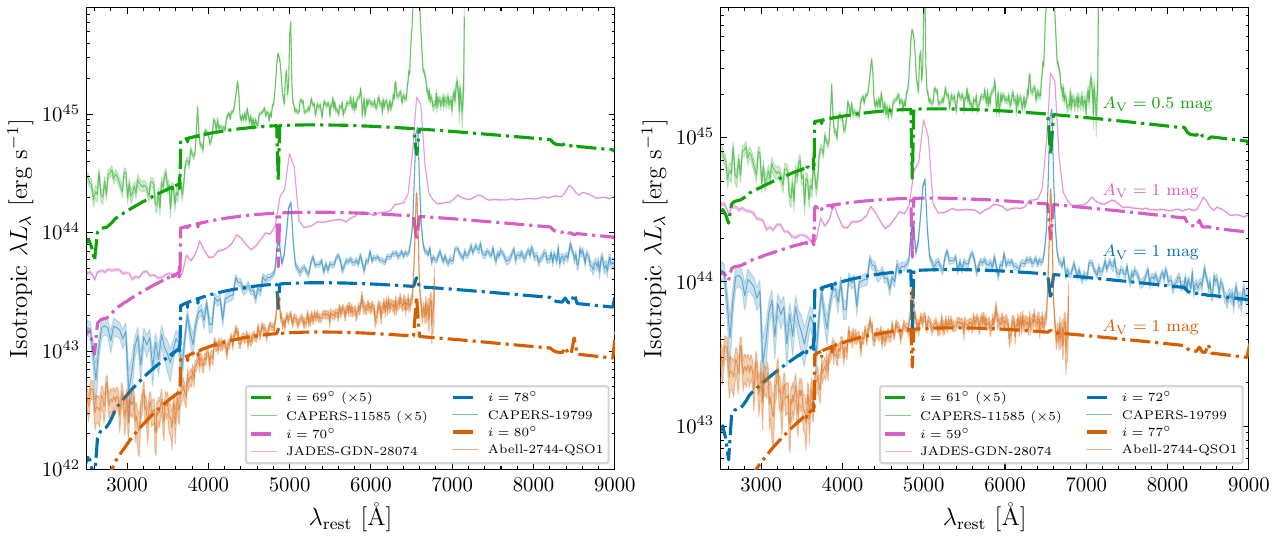}
\caption{Qualitative comparison of observations and simulations. The dashed curves show model SEDs for an SMBH with $M_{\mathrm{BH}}= 10^7 M_{\odot}$ and $\dot{m}=1000$ at various inclination angles. The solid curves and shaded regions represent observations with their $1\sigma$ uncertainties. In the right panel, we apply the \citet{Calzetti2000} dust attenuation law, with varying extinction coefficients, to ``deredden'' the observations. Observations with low dust extinction roughly match our model predictions. \label{fig5-obs}}
\end{figure*}

We present a qualitative comparison between simulated emergent SEDs and observations in Figure~\ref{fig5-obs}. Based on the LRD-to-total flux fraction estimations in \citet{Sun2026}, we randomly select four sources with the LRD flux fraction at $5500\ \text{\AA}$ larger than $80\%$. CAPERS 11585 and 19799 were observed as part of the CAPERS survey (GO-6368; PI: M. Dickinson). JADES-GND-28074, one of the most well-characterized nearby LRDs at $z=2.26$ \citep{Juodzbalis2024}, was observed by the JADES program \citep[GO-1811;][]{Rieke2023_jades, Bunker2024_jades, Eugenio2025_jades, Eisenstein2026_jades}. The NIRSpec/PRISM spectroscopic data for these three sources are retrieved from the DAWN JWST Archive \citep[version 4.4\footnote{\url{https://dawn-cph.github.io/dja/spectroscopy/nirspec/}};][]{Brammer2025_dawn}, with data processing described in \citet{de_Graaff2025_dawn} and \citet{Heintz2025}, and incorporating the v4 improvements detailed by \citet{Valentino2025_dawn} and \citet{Pollock2026}. Additionally, we include Abell-2744-QSO1, a strongly lensed LRD at $z=7.04$ \citep[e.g.,][]{Furtak2024, JiXH2025a}. The spectrum presented here is derived from image A \citep{JiXH2025a}, which is reduced using NIRSpec/PRISM Micro-Shutter Assembly data and the GTO pipeline reduction method described in \citet{Scholtz2025_pipline}. To account for the gravitational lensing effect of this source, we apply a magnification correction factor of $\mu =6$ \citep{JiXH2025a}. The model-to-data comparison presented here should be regarded as a demonstration rather than a rigorous SED fit (i.e., without fine-tuning). As shown in the left panel of Figure~\ref{fig5-obs}, the emergent SEDs of an SMBH with $M_{\mathrm{BH}}= 10^7 M_{\odot}$ and $\dot{m}=1000$ at high inclinations are broadly consistent with observed red continua and Balmer breaks. To further improve the agreement between the model and the observed data, we incorporate the \citet{Calzetti2000} attenuation law (right panel of Figure~\ref{fig5-obs}) with a modest amount of dust extinction of $A_{V}\lesssim 1.5$ mag. 

\begin{figure*}
\centering
\includegraphics[width=.8\linewidth]{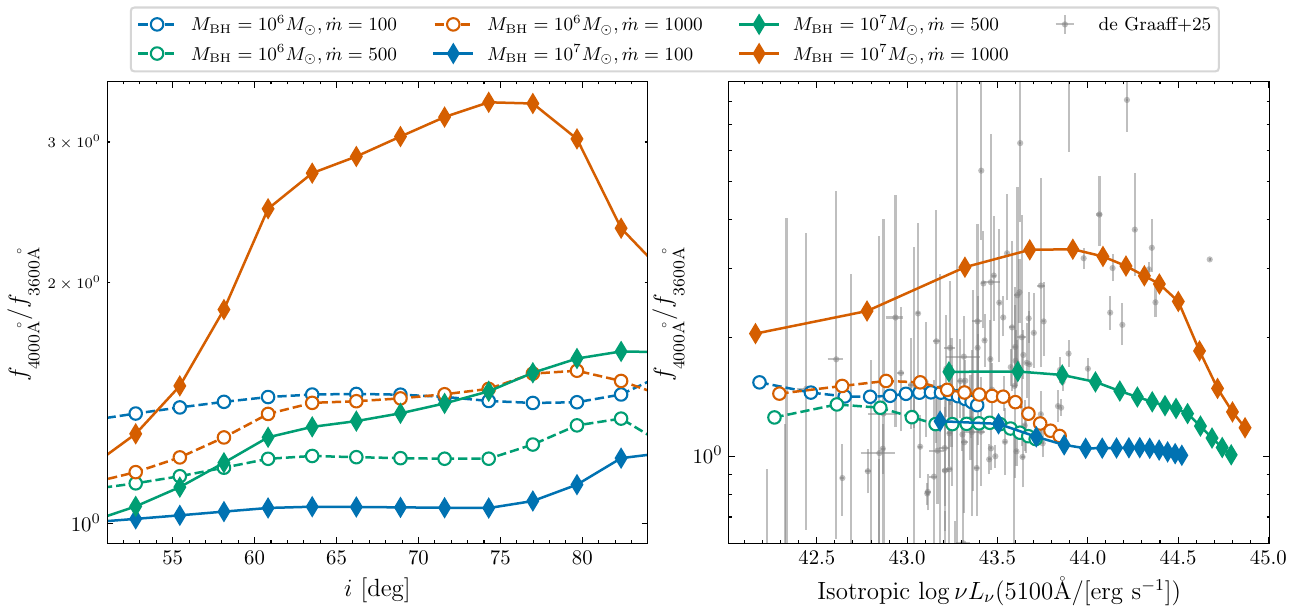}
\caption{Balmer break strength ($f_\mathrm{4000\text{\AA}}/f_\mathrm{3600\text{\AA}}$) as a function of inclination angle and monochromatic luminosity at $5100\ \text{\AA}$. Open circles and filled diamonds represent SMBHs with  $M_{\mathrm{BH}}= 10^6 M_{\odot}$ and $10^7 M_{\odot}$, respectively. Blue, green, and orange markers indicate accretion rates $\dot{m}= 100$, $500$, and $1000$, respectively. In the right panel, gray dots denote observational data from \citet{deGraaff2025}, with error bars representing $1\sigma$ uncertainties. The Balmer break strength generally increases with inclination angle, and the observational data are broadly consistent with our simulations.
\label{fig6-break}}
\end{figure*}

Prominent Balmer breaks that cannot be explained by the typical stellar population have been detected in some LRDs \citep[e.g.,][]{JiXH2025a, deGraaff2025, Naidu2025}, providing important clues about the physical conditions in emission regions. We quantify the Balmer break strengths of simulated emergent SEDs with different black hole masses and Eddington accretion ratios and compare them with observations in Figure~\ref{fig6-break}. While \citet{deGraaff2025} define the Balmer break strength using the flux ratio between two rest-frame wavelength windows ($[3620,3720]\ \text{\AA}$ and $[4000,4100]\ \text{\AA}$), we adopt a monochromatic flux ratio $f_\mathrm{4000\text{\AA}}/f_\mathrm{3600\text{\AA}}$ between the line-free wavelength $4000\ \text{\AA}$ and $3600\ \text{\AA}$ to characterize the Balmer break strength of simulated SEDs. As shown in the left panel of Figure~\ref{fig6-break}, the Balmer break strength generally increases with inclination angles. This trend is caused by the fact that higher inclinations provide a more obscured view of the hot inner disk, allowing the cool ($T\lesssim10^4\ \mathrm{K}$), dense outer regions to dominate the emergent spectrum. The right panel of Figure~\ref{fig6-break} shows the break strength as a function of the monochromatic luminosity at $5100\ \text{\AA}$. Observations from \citet{deGraaff2025} are broadly consistent with our hyper-Eddington accretion model predictions. In addition, we test the impact of the viscosity parameter $\alpha$ on the Balmer break strength. We find that adopting a smaller viscosity parameter ($\alpha = 0.01$) modifies the vertical structure of the disk, resulting in a more pronounced Balmer break strength than with $\alpha = 0.1$ (see Appendix~\ref{appendix:break}).

\subsection{The broad emission lines}\label{sect:bel}
In our hyper-Eddington accretion model, BLR clouds in the polar regions are photoionized by UV photons from the inner SLIM disk, producing the observed BELs (panel d in Figure~\ref{fig1-model}). This geometry naturally produces the observed broad Ly$\alpha$ \citep[e.g.,][]{Ji2026_lya} through ionizing photons in low-inclination directions. While a large number of BLR clouds may form near the equatorial plane, the ionizing radiation field is highly anisotropic due to the significant self-shielding of the puffed-up SLIM disk: polar clouds receive substantially more UV photons than equatorial ones. Hence, the polar clouds that are most efficient at producing BELs should also be located farther away than equatorial clouds, which is consistent with BLR stratification. In the simplest case, the BEL profile can be described as the superposition of high- and low-velocity Gaussian components, which together resemble a Lorentzian profile or an exponential profile \citep[e.g.,][]{Wang2014, Scholtz2026, Madau2026_BLR}. In the local universe, narrow-line Seyfert 1 galaxies, some of which are likely powered by super-Eddington accretion onto relatively small SMBHs, often show Lorentzian BEL profiles \citep[e.g.,][]{Zhou2006, Berton2020} and can be understood in the same framework \citep{Wang2014}. Interestingly, a large fraction of LRDs are well characterized by Lorentzian or two-Gaussian profiles. While such non-Gaussian features have recently been interpreted as signatures of electron scattering in dense gas \citep[e.g.,][]{Chang2026, Rusakov2026}, we argue that these profiles arise naturally from BLRs in super-Eddington systems  \citep[see also][]{Brazzini2025, Brazzini2026, Scholtz2026, Madau2026_BLR}. Future spectropolarimetric observations could potentially distinguish between these mechanisms because electron scattering is expected to produce significant polarization signatures in continua and lines.

If the BLR clouds follow a disk-like Keplerian velocity field, a high inclination angle will naturally correspond to a large line-of-sight velocity, resulting in broader lines. Our model therefore predicts a positive correlation between Balmer break strength and the full width at half maximum (FWHM) of the BELs, as tentatively observed by \cite{Matthee2026}. Consequently, single-epoch black hole mass estimators calibrated on low-inclination type-1 AGNs may introduce an inclination-dependent bias when applied to LRDs. Assuming a mean inclination of $30^\circ$ for type-1 AGNs and $75^\circ$ for LRDs, the inclination effect induces an overestimation factor of $(\sin 75^\circ/\sin 30^\circ)^2 \approx 3.7$. 

We also briefly discuss the equivalent widths (EWs) of BELs. The observed EW traces the ratio of the line flux to the local continuum flux. The line flux is proportional to the number of ionizing photons intercepted by the BLR clouds, while the observed continuum is highly anisotropic due to the geometric self-shielding of the puffed-up SLIM disk. Consequently, at large inclination angles, the observed optical continuum flux is suppressed, which is consistent with the observed optical luminosity falling below local relation expectations \citep{Chen2026_ABCD}, ultimately resulting in an increased EW of $\text{H}\alpha$. Due to the anisotropic ionizing structure of BLRs and the intrinsic deficiency of high-energy ionizing photons in hyper-Eddington accretion disks, high-ionization lines such as broad \HeII\ in LRDs and LBDs are expected to be weak or absent. These effects have been well demonstrated by \citet{Madau2026} under the ``Polish doughnut'' framework and should also apply in our model (for the differences between the ``Polish doughnut'' and our model are discussed in Section~\ref{sect:dis}). Quantitatively reproducing the detailed line shapes, equivalent widths, and Balmer decrements highly depends on the kinematic and photoionization modeling of the BLR clouds. Comprehensive BLR modeling within the hyper-Eddington framework is the focus of our future work. 

\subsection{The optical variability}\label{sect:var}
In our hyper-Eddington accretion model, the red optical continuum of LRDs is produced by the self-shielded disk at the typical radius of $10^4\ R_{\mathrm{S}}$. Due to the large accretion rate and radius, the local thermal timescale ($t_\mathrm{th}\sim \alpha ^{-1}(R_\mathrm{adv}/R_\mathrm{S})^{3/2}GM_\mathrm{BH}c^{-3}$) for $\dot{m}=1000$ is approximately $30(M_{\mathrm{BH}}/[10^7M_{\odot}])$ years in the rest frame \citep{Zhou2024a, Zhou2025}. This duration is much longer than the timescales of current LRD monitoring campaigns (ranging from months to a few years) and is roughly consistent with the rest-frame variability timescale ($\sim 25$ years) inferred from a gravitationally lensed LRD by \citet{ZJZhang2025}. In addition, the hyper-Eddington accretion process is also expected to suppress the variability amplitudes \citep[e.g.,][]{ZJZhang2025-var, Secunda2026}. In our model, the lack of month-long variability in LRDs tentatively suggests that they are likely to host SMBHs with $M_\mathrm{BH}\gg 10^5 M_\odot$. 

\begin{figure}
\centering
\includegraphics[width=1.\linewidth]{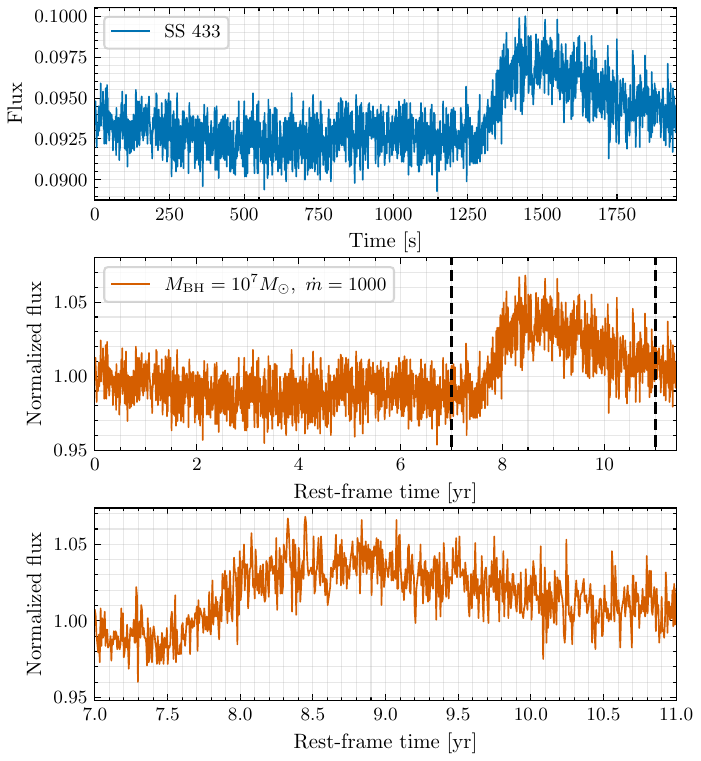}
\caption{Mocked optical light curve of an LRD ($M_{\mathrm{BH}}= 10^7 M_{\odot}$ and $\dot{m}=1000$), derived by scaling the characteristic thermal timescale of SS~433. The upper panel displays the $R$-band light curve of SS~433 during a nearly edge-on view, with a median sampling interval of $\sim 1\ \mathrm{s}$ obtained by \citep{Burenin2011}. The middle panel presents the simulated LRD optical light curve, scaled by the ratio of the thermal timescales at the maximum temperature radius (Figure~\ref{fig2-teff}), i.e., $f_\mathrm{scale}=t_\mathrm{th,LRD}/t_\mathrm{th,SS~433}$. The bottom panel provides a zoomed view of the intervals indicated by the vertical black dashed lines in the middle panel. Assuming LRDs and SS~433 share a common variability mechanism, we predict that LRDs will exhibit variability over several years in the rest frame.
\label{fig7-lc}}
\end{figure}

To qualitatively predict the long-term optical variability of LRDs, we generate a mock light curve based on the hypothesis that LRDs and SS~433 share a similar variability mechanism. We use the $R$-band light curve of SS~433 when the accretion disk is viewed edge-on as our baseline (upper panel of Figure~\ref{fig7-lc}). The $R$-band data are obtained from the Russian Turkish 1.5-m telescope with a median sampling interval of $\sim 1\ \mathrm{s}$ \citep{Burenin2011}. By matching the observed characteristic temperature of SS~433 \citep[$5\times 10^{4}$--$7\times 10^{4}$ K;][]{Dolan1997} to our estimated maximum observable temperatures derived in Section~\ref{sect:cont} (Figure~\ref{fig2-teff}), we infer a dimensionless accretion ratio of $\dot{m}\approx626$ with fixing the black hole mass of $M_\mathrm{BH}\approx6\ M_\odot$. We then scale the time axis of the SS~433 light curve onto an LRD. The timescale scaling factor $f_\mathrm{scale}=t_\mathrm{th,LRD}/t_\mathrm{th,SS~433}$ is defined as the ratio of the thermal timescales at the maximum temperature radius for both systems. The middle panel of Figure~\ref{fig7-lc} shows the scaled light curve of an LRD with $M_{\mathrm{BH}}= 10^7 M_{\odot}$ and $\dot{m}=1000$. Hence, the variability timescale for the LRD is on the order of several years (rest frame). 

UV emission from the innermost disk regions photoionizes the BLR clouds, producing the observed BELs. Within the standard reverberation-mapping framework, the BELs should respond to variability in the ionizing continuum. For an SMBH with $M_{\mathrm{BH}}= 10^7 M_{\odot}$ and $\dot{m}=1000$, the face-on luminosity of $\nu L_\mathrm{\nu}(5100\mathrm{\AA}) \approx 10^{45}\ \mathrm{erg\ s^{-1}}$ implies a BLR light-travel time of $\sim 180$ days according to the $R$--$L$ relation \citep{Bentz2013}. Although light-travel smoothing across the extended BLR suppresses the amplitude of BEL variability, its variation timescales remain significantly shorter than the thermal timescale of the optical continuum. Hence, we expect the variability of BELs to be easier to detect than that of the red optical continuum. Variations in the EWs of H$\alpha$ have been observed in the local LRD `Lord of LRDs', whereas there is no photometric variability beyond measurement uncertainties \citep[e.g.,][]{JiXH2026, Lambrides2026}. A narrow filter centered on the BELs can be used to detect BEL variability and probe the variability properties of LRDs. 

In our framework, LBDs are either the same class of hyper-Eddington SMBHs seen at lower inclination or super-Eddington SMBHs with lower $\dot{m}$ (panels c and d in Figure~\ref{fig1-model}). Therefore, since the inner disk regions are visible to the observer in these cases, LBDs should exhibit more pronounced UV/optical variability than LRDs. 

If the disk scale height changes due to fluctuations in the accretion rate, one would expect a transition from an LRD to an LBD, and vice versa. The characteristic timescale for the scale height to change around the truncation radius would be the dynamical timescale, which is about $3(M_{\mathrm{BH}}/[10^7\ M_{\odot}])$ years in the rest frame.

\subsection{The apparent Eddington ratio}\label{sect:ledd}
Due to geometric self-shielding, the observed SED of an LRD is much softer than the intrinsic one. Thus, the bolometric luminosity inferred from the observed SED can be much lower than the true bolometric luminosity, leading to a substantial underestimation of the Eddington ratio $L_{\mathrm{bol}}/L_{\mathrm{Edd}}$. At high inclinations ($60^\circ \lesssim i \lesssim 80^\circ$), the observed isotropic (i.e., incorrectly assume the radiation is isotropic) continuum luminosity for $M_{\mathrm{BH}}=10^7\ M_{\odot}$ and $\dot{m}=1000$ at rest-frame $5100\ \anst$ is $2\times 10^{43} \lesssim \nu L_\mathrm{\nu}(5100\mathrm{\AA})/\mathrm{[erg\ s^{-1}]} \lesssim 3\times 10^{44}$ (Figure~\ref{fig4-spec}). If we adopt a bolometric correction factor of $10$ for the luminosity at rest-frame $5100\ \anst$, the apparent Eddington ratio is $0.16$--$3$ between $60^\circ \lesssim i \lesssim 80^\circ$. Furthermore, if the high inclination also leads to an overestimation of $M_{\mathrm{BH}}$, then the apparent Eddington ratio would be further suppressed. In our model, hyper-Eddington LRDs can therefore appear as sub-Eddington accretion. 

Observational studies often use the broad H$\alpha$ line luminosity to infer the bolometric luminosity \citep[e.g.,][]{Maiolino2024, JiXH2025a, Juodzbalis2026-bol}; this conversion should be more reliable than the $5100\ \anst$ proxy but depends on the BLR covering factor. In our model, the BLRs of LRDs may differ from those in normal AGNs because the ionizing emission of the latter is more isotropic. Therefore, the conversion of the broad H$\alpha$ line to the bolometric luminosity might be underestimated for LRDs or other super-Eddington sources if the BLR clouds tend to be equatorially distributed. Indeed, there is evidence that the H$\alpha$-based bolometric luminosity underestimates the actual one \citep[see, e.g.,][]{Tortosa2026} for local super-Eddington accreting massive black holes \citep[e.g.,][]{Du2014}. In addition, the possible overestimation of the virial mass mentioned in Section~\ref{sect:bel} also reduces the apparent Eddington ratio.

\subsection{The radio emission}\label{sect:radio}
LRDs are often found to be radio-undetected \citep[e.g.,][]{Perger2025, Mazzolari2026}. In radio-loud AGNs, the radio emission is considered to originate from relativistic jets launched from the innermost regions of the accretion disk. In the hyper-Eddington accretion case, the jet from the innermost SLIM disk can be baryonic, resulting in a lower velocity due to the high accretion rate. Indeed, the jet in SS~433 has a velocity of $\sim 0.26c$ \citep{Fabrika2007}; the jet in M81 ULS-1 has a velocity of $\sim 0.17c$ \citep{Liu2015}. Radio emission in M81 ULS-1 is undetected, possibly due to its extragalactic distance \citep{Liu2015}. The radio flux of SS~433 is around $1$ Jy at a distance of $5$ kpc \citep{Fabrika2007}. If we scale the radio flux of SS~433 by the black hole mass ratio and the square of the luminosity distance ratio, the expected radio flux for an LRD with $M_{\mathrm{BH}}=10^7\ M_{\odot}$ at $z=0.2$ ($z=6$) is around $20\ \mathrm{\mu Jy}$ ($7\ \mathrm{nJy}$). These estimates suggest that radio signatures from local LRD analogs should be detectable through deep Very Large Array (VLA) exposures. Indeed, radio emission is detected in the local LRD analog SDSS J104755.92+073951.2, with a radio flux of $\simeq 100\ \mathrm{\mu Jy}$ at $6.0$ GHz \citep{Rodriguez2026}. Meanwhile, radio emission from a high-redshift LRD may be detectable only with the Square Kilometer Array (SKA) or the FAST Core Array \citep{Jiang2024}. 

Interactions between powerful baryonic jets and the ambient medium can drive shocks and generate non-thermal high-energy emission, as observed in the SS 433/W50 system \citep[e.g.,][]{Abeysekara2018, Bykov2025}. Similar jet–ambient-medium interactions in LRDs may therefore produce high-energy $\gamma$-ray emission. As an order-of-magnitude estimate, scaling the observed $>1$ GeV $\gamma$-ray flux of SS 433 \citep{Fang2020} by the black hole mass ratio and the inverse square of the luminosity-distance ratio yields an expected photon flux of $\sim10^{-18}\ {\rm ph\ cm^{-2}\ s^{-1}}$ for an LRD hosting a $10^{7}\,M_\odot$ black hole at $z=6$. This flux is far below the sensitivity of current $\gamma$-ray facilities (which is of order $10^{-10}\ {\rm ph\ cm^{-2}\ s^{-1}}$ for Fermi-LAT\footnote{\url{https://fermi.gsfc.nasa.gov/ssc/data/analysis/documentation/Cicerone/Cicerone_LAT_IRFs/LAT_sensitivity.html}} at $1$ GeV) and is therefore not detectable unless $z\lesssim 0.001$. Moreover, baryonic jets may produce neutrinos through hadronic interactions. Therefore, neutrinos would provide a unique probe of hyper-Eddington accretion in LRDs because they can propagate through dense gas. Due to their high number density, LRDs have even been suggested as significant contributors to the diffuse neutrino background \citep[e.g.,][]{Kuze2026}. Future observations by neutrino telescopes could verify this idea.

\begin{deluxetable}{llllll}
\label{tab:model}
\tabletypesize{\footnotesize}
\tablewidth{0pt}
\tablecaption{SMBH accretion models}
\renewcommand{\arraystretch}{2.}
\tablehead{
\colhead{Model} & \colhead{Gas-enshrouded AGN} & \colhead{BLR-attenuation} & \colhead{Polish donut} & \colhead{\shortstack{Spherical super-\\Eddington accretion}} & \colhead{\shortstack{Hyper-Eddington\\SLIM disk}}
}
\startdata
\parbox[c]{2.2cm}{Reference} & 
\parbox[c]{2.5cm}{e.g., \citet{Kido2025, Naidu2025, Begelman2026, Rusakov2026, Sneppen2026}} & 
\parbox[c]{2.5cm}{\citet{Inayoshi2025, JiXH2025a}} & 
\parbox[c]{2.5cm}{\citet{Madau2026, Madau2026_BLR}} & 
\parbox[c]{2.4cm}{\citet{LiuHP2025}} &
This work \\
\tableline
\parbox[c]{2.2cm}{Central engine}  &
\parbox[c]{2.5cm}{Black-hole star / quasi-star / non-spherical cocoons}&
\parbox[c]{2.5cm}{AGN with dense gas clumps} &
\parbox[c]{2.5cm}{Super-Eddington non-advective accretion (a.k.a., Polish donut)} &
\parbox[c]{2.4cm}{Super-Eddington spherical accretion} &
\parbox[c]{2.5cm}{Hyper-Eddington accretion (advection dominated)}\\
\tableline
\parbox[c]{2.2cm}{Red optical continuum} &
\parbox[c]{2.5cm}{Fully-covered dense gas envelop} &
\parbox[c]{2.5cm}{AGN continuum attenuated by BLR-associated dense gas + Dust reddening} &
\parbox[c]{2.5cm}{Polish donut continuum attenuated by BLR-associated dense gas + Dust reddening} &
\parbox[c]{2.4cm}{Spherical photosphere} &
\parbox[c]{2.5cm}{Self-shielded disk photosphere (weak variability)} \\
\tableline 
\parbox[c]{2.2cm}{UV continuum} &
\parbox[c]{2.5cm}{Stellar population} &
\parbox[c]{2.5cm}{Attenuated AGN continuum/Stellar population} &
\parbox[c]{2.5cm}{Polish donut continuum attenuated by BLR-associated dense gas + Dust reddening} &
--- &
\parbox[c]{2.5cm}{Nebular emission (non-variable) or scattered AGN emission (variable) or stellar population?}\\
\tableline
\parbox[c]{2.2cm}{Ionization continuum} &
{AGN} &
AGN & \parbox[c]{2.5cm}{Polish donut anisotropic emission} &
\parbox[c]{2.5cm}{---} &
\parbox[c]{2.5cm}{Inner anisotropic emission} \\
\tableline
\parbox[c]{2.2cm}{Balmer break} &
\parbox[c]{2.5cm}{Fully-covered dense gas envelop} &
\parbox[c]{2.5cm}{BLR-associated dense gas} &
\parbox[c]{2.5cm}{BLR-associated dense gas} &
\parbox[c]{2.5cm}{Spherical photosphere} &
\parbox[c]{2.5cm}{Self-shielded disk photosphere}\\
\tableline 
\parbox[c]{2.2cm}{Balmer absorption lines} &
\parbox[c]{2.5cm}{Fully-covered dense gas envelop} &
\parbox[c]{2.5cm}{BLR-associated dense gas} &
\parbox[c]{2.5cm}{BLR-associated dense gas/torus} &
--- &
\parbox[c]{2.5cm}{BLR or circumnuclear clouds?}\\
\tableline 
\parbox[c]{2.2cm}{Emission line profile} &
\parbox[c]{2.5cm}{Scattering} &
--- &
\parbox[c]{2.5cm}{Stratified BLR clouds} &
--- &
\parbox[c]{2.5cm}{Stratified BLR clouds}\\
\tableline
\parbox[c]{2.2cm}{X-ray weakness} &
\parbox[c]{2.5cm}{Fully-covered dense gas envelop} &
\parbox[c]{2.5cm}{BLR-associated dense gas obscuration} &
\parbox[c]{2.5cm}{Intrinsically weak + Self-shielding} &
--- &
\parbox[c]{2.5cm}{Intrinsically weak + Self-shielding}\\
\tableline
\parbox[c]{2.2cm}{Radio emssion} &
\parbox[c]{2.5cm}{Fully-covered dense gas envelop blocks radio emission} &
--- & --- & --- &
Baryonic jet \\
\enddata
\end{deluxetable}

\section{Discussion}\label{sect:dis}
Motivated by the observational features (softer-than-expected SED and anisotropic nature) of the hyper-Eddington X-ray binary SS~433, we propose that LRDs are hyper-Eddington accretion SMBHs ($\dot{m}\gtrsim 500$) at high inclinations. The model naturally explains the red optical continuum, the prominent Balmer break, the observed BEL properties, the X-ray weakness, the weak variability, and the apparent sub-Eddington luminosities. The key ingredient is geometric self-shielding by the puffed-up SLIM disk, which obscures the high-energy emission from the inner regions and makes the emergent SED highly inclination-dependent. This geometry naturally links LRDs and (part of) the LBDs as high- and low-inclination counterparts of the same underlying accretion structure. 

The origin of the UV component remains uncertain: it may be scattered inner-disk light, nebular emission due to the interaction between an SS 433-like slow jet and the ambient gas, or emission from a young stellar population. Moreover, inner winds, which are likely to be present, may be optically thick. They could therefore soften the emission from the inner disk and contribute to the observed UV emission. Because the gas opacity peaks at temperatures of $10^4\sim 10^5\ \mathrm{K}$, the radiation-pressure-driven optically thick winds may efficiently absorb and reprocess the disk radiation as a $10^4\sim 10^5\ \mathrm{K}$ blackbody. This is consistent with the wind framework expectation of \citet{Soria2016} for the SMBH cases. If dominated by scattered light, the UV continuum should be significantly polarized, with LRDs and LBDs showing similar polarized spectral signatures at a given accretion rate. The V-shaped SED can then be reproduced because the disk emission cannot extend to UV.

\subsection{Comparison to other models}\label{sect:comp}
Several SMBH accretion models have been proposed to explain the nature of LRDs (see Table~\ref{tab:model}). For example, \cite{Madau2026} propose that LRDs are powered by super-Eddington accretion onto SMBHs with large inclination angles; our model shares many similarities with their model. In their model, the X-ray weakness can also be explained naturally \citep[also see, e.g.,][]{Pacucci2024}. One major difference is that their model is based on the ``Polish doughnut'' model \citep[e.g.,][]{PB1980}, whereas ours is based on a SLIM disk. While the SLIM disk is advection-dominated, the ``Polish doughnut'' is constructed using a static geometric model with a presumed specific angular momentum distribution at its photosphere, rather than solving the disk equations. The scale height of the ``Polish doughnut'' can be much larger than $R$. The vertical structure of the ``Polish doughnut'' is often neglected, making it difficult to assess whether the structure itself can reproduce the observed Balmer break. In \citet{Madau2026}, the Balmer break is attributed to absorption by the dense BLR gas, whereas in our model it is produced by radiative transfer through the SLIM-disk atmosphere. In addition, they attribute the V-shaped SED across both the UV and optical bands to the joint effects of self-shielding and dust extinction. Their model requires a relatively large visual extinction \citep[$A_V \sim 2.8$ mag;][]{Madau2026_2}, in contrast to the modest extinction required in our framework ($A_V \lesssim 1$ mag).

The ``Polish doughnut'' model of \cite{Madau2026} and our model have different predictions for continuum and line variability. First, the BLR-attenuation scenario \citep[e.g.,][]{Inayoshi2025, JiXH2025a, Madau2026} can be tested through the cross-correlation of variability of the BELs and the red optical continuum. If BLR attenuation is responsible for the red optical continuum, the BELs should show the ``holiday effect'' (i.e., the BELs do not respond to the variability of the optical continuum), as observed in NGC~5548 \citep{Goad2016, Dehghanian2019}. Meanwhile, the optical continuum variability may be more prominent in the ``Polish doughnut'' model than in our model. If the gravitational energy released in the ``Polish doughnut'' changes, its geometric structure should adjust on the dynamical timescale of $\sim 0.15 (M_{\mathrm{BH}}/(10^7\ M_{\odot}))(R/(1000 R_{\mathrm{S}}))^{1.5}$ years; then, the optical continuum would be expected to vary on such timescales. In the meantime, the optical variability timescale of our model is about $30$ years (Section~\ref{sect:var}). Future time-domain studies of LRDs can distinguish the two models.

Another highly relevant model is the super-Eddington disk model presented by \cite{LiuHP2025}, which is based on either a standard SSD or spherical accretion.\footnote{When submitting this manuscript, we are aware of the spherical wind model by \citet{LiuJR2026}, which (in our point of view) shares many similarities with the spherical accretion model of \citet{LiuHP2025} but with an opposite velocity and thus will not be discussed separately.} The authors presented excellent radiative calculations demonstrating that the Balmer break and red continuum can be produced intrinsically by super-Eddington accretion, without invoking gas or dust attenuation. They did not consider the self-shielding effect and the anisotropic nature of the SED, which are key to explaining the observed SEDs and BELs of LRDs. They also did not mention the connection between LRDs and ULSs, which motivates this work. 

Much previous work has focused on the black-hole star \citep[e.g.,][]{Kido2025} or quasi-star \citep{Begelman2008,Begelman2026} scenario. In these scenarios, an SMBH is surrounded by a large amount of gas with a temperature of $\sim 5000$ K \citep[as predicted by][]{Begelman2008}, which can produce the red optical continuum \citep[e.g.,][]{deGraaff2025} and the Balmer break \citep[e.g.,][]{Naidu2025, Gentile2026}. The main differences between these scenarios and our model are as follows. First, the red optical continuum and the Balmer break in the black-hole star or quasi-star scenario are produced by the large amount of gas surrounding the SMBH, whereas those in our model are produced by the self-shielded SLIM disk. Second, the black-hole star or quasi-star scenario is based on spherical accretion, whereas our model is built upon disk hyper-Eddington accretion, which shows anisotropic emission. Third, the quasi-star scenario always predicts scatter-broadened BELs within the gas envelope, whereas in our model BELs are emitted by BLRs. Hence, if the optical continuum and the BLEs originate from the same gas envelope in the quasi-star scenario, their polarizations should be similar; for our model, while the optical continuum is expected to be polarized due to electron scattering in the disk atmosphere, the BELs of LRDs are produced in the separate BLRs (just as in normal type I AGNs) and could have different polarization levels and directions. Future spectropolarimetric observations \citep[for a recent example, see][]{DEugenio2026} of BELs would be crucial in distinguishing our model from the black-hole star/quasi-star model.

Several works have also proposed that the V-shaped SEDs of LRDs are due to the joint contribution of an inner accretion disk and additional stellar heating in the outer self-gravitating regions. For example, \cite{Zhang2026} assume that the inner accretion disk is a standard SSD and that the outer self-gravity regions are heated by star formation. Meanwhile, \cite{Wang2025b} consider the inner accretion disk to be a SLIM disk and the outer self-gravity regions to be heated by stellar-mass black hole accretion, which is motivated by disk-size studies \citep{Zhou2024b, Wang2025a}. Very recently, \cite{Chen2026} argued that stellar objects in the outer self-gravity regions can hollow out the inner disk and that its emission is responsible for the observed universal effective temperature of $\sim 5000$ K; instead, the UV emission is produced by stellar populations on galaxy scales. Hence, we conclude that all three models predict extremely weak optical-continuum variability because of statistical averaging effects. The UV emission can be variable in the models of \cite{Zhang2026} and \cite{Wang2025b}, with the former expected to be more variable than the latter. The BELs can vary in the models of \cite{Zhang2026} and \cite{Wang2025b} due to variability in the ionizing UV continuum, but not in the model of \cite{Chen2026} due to the lack of disk UV photons. Hence, a future time-domain joint study of the UV/optical continuum and BELs can be used to test these models and ours. We also note that additional stellar heating scenarios can be combined with our model, which may be necessary to explain the observed SEDs of some LRDs.

\subsection{SMBH mass growth}\label{sect:massgrowth}
If LRDs are indeed powered by hyper-Eddington accretion with an inflowing accretion rate of $\dot{m}\gtrsim 500$ and large inclination angles, this has important implications for SMBH mass growth in the early universe. However, due to the large inclination angles, we are unable to directly probe the inner structure of the SLIM disk and the actual accretion rate at the SMBH event horizon. The BELs from BLR clouds in the polar regions might be used to probe the ionizing continuum from the innermost regions. The inferred results would depend on the BLR geometry, which is still unclear. Reverberation-mapping observations might be of limited help because the observed optical continuum is unlikely to be a good proxy for the ionizing continuum. Spatial resolution of BLR clouds is a promising way to study them. Current GRAVITY+ observations \citep{GravityPlus2022} are already making progress, and next-generation interferometric facilities could provide the necessary resolution and sensitivity to resolve these distant systems. 

Within our hyper-Eddington accretion scenario, we discuss SMBH mass growth during the LRD phase by considering two aspects. First, as we argued in Section~\ref{sect:framework}, the hot inner regions with  $T \gtrsim 10^4$ K can launch powerful, radiation-pressure-driven winds \citep[e.g.,][]{SSD, King2003, Cao2022} that are strongly decelerated in the cold outer regions, where the opacities become extremely small, and subsequently rejoin the accretion flow. Consequently, while the mass accretion rate in the outer regions remains high, its value in the inner regions is greatly reduced; the net accretion rate is also far smaller than the accretion rate in the outer regions. Nevertheless, even if the effective inflow at the event horizon is only $\dot{m}\sim 10$, the corresponding e-folding timescale is $\sim 0.1\,t_{\mathrm{ST}}$, where $t_{\mathrm{ST}}\simeq 5\times 10^7$ yr, i.e., $\sim 5\times 10^6$ yr. This rapid growth can accommodate the large SMBH masses inferred at $z\sim 7$, and it naturally helps explain the high $M_{\mathrm{BH}}/M_\star$ ratios and overmassive black holes seen in LRD populations \citep[e.g.,][]{Juodzbalis2025-mass}. 

Second, the duration of the LRD phase (i.e., the duty cycle) also constrains the total mass accumulated during the hyper-Eddington stage. Previous studies have estimated a duty cycle of $\sim 1\%$ by comparing the UV luminosity functions (LFs) of LRDs and galaxies \citep[e.g.,][]{Sun2026}. However, this estimation implicitly assumes that the rest-frame UV emission of LRDs originates entirely from their host galaxies. If the UV continuum of LRDs contains a non-negligible contribution from the AGN itself (e.g., scattered emission) or nebular emission \citep[e.g.,][]{Labbe2024, CHChen2025}, then the intrinsic host galaxy UV luminosity would be overestimated since the galaxy number density should increase with decreasing luminosity. This would lead to an overestimation of the duty cycle. A reduced duty cycle directly limits the total mass accreted during the hyper-Eddington LRD phase.

\subsection{Caveats}
It is important to acknowledge that the vertical structure of the hyper-Eddington accretion disk, as derived in Section~\ref{sect:balmer}, omits certain physical complexities, such as convection, winds, self-illumination, and other effects that could alter the disk structures in hyper-Eddington accretion regimes. To evaluate the impact of these uncertainties on our primary results, we tested the sensitivity of the Balmer break to the vertical gas density distribution. We compare the Balmer break strength derived from our vertical density profile with that derived from a simplified case assuming a uniform gas density across the vertical dimension (see Appendix~\ref{appendix:break}). We find that the Balmer break remains prominent even under the simplified assumption of uniform density. This comparison shows that, although our current vertical-structure calculations may not capture all the physical details, the prediction of strong Balmer breaks is robust.

There are several simplifications in our radiative transfer calculations that should be noted. First, our calculations assume LTE. While this is a reasonable approximation for dense, optically thick regions, lower-density upper layers may exhibit non-LTE effects. These deviations could alter the ionization states and level populations of hydrogen, potentially modifying emission/absorption lines in the emergent SED. Second, our radiative transfer calculation is static and does not consider the gas kinematics within the accretion disk, which could affect the break and line profiles. 

Our current discussion of the BLR is limited to a simplified geometric and qualitative kinematic description. We have not yet performed detailed kinematic and photoionization modeling of BLR clouds illuminated by the highly anisotropic continuum of the inner SLIM disk. Consequently, the precise line profiles, equivalent widths, and Balmer decrements, as well as their impact on bolometric luminosity estimations, remain to be fully quantified. Comprehensive modeling of the BLR within this hyper-Eddington framework will be the primary focus of our future work. 

\section{Summary}\label{sect:sum}
Motivated by the softer-than-expected SED and anisotropic nature of SS~433, we propose a hyper-Eddington SMBH accretion model with $\dot{m}\gtrsim 500$ to explain the observational features of LRDs. Our results can be summarized as follows. 
\begin{itemize}
    \item The red optical continuum, with characteristic effective temperature $\sim 5000$ K, can arise naturally from self-shielding in the SLIM disk. We therefore interpret LRDs as high-inclination, hyper-Eddington SMBHs, with LBDs representing the low-inclination or relatively low-accretion-rate counterparts (Section~\ref{sect:cont}; Figures~\ref{fig1-model} and \ref{fig2-teff}). 
    \item The Balmer break is produced by radiative transfer in the outer disk photosphere (Section~\ref{sect:balmer}; Figures~\ref{fig3-vs} and \ref{fig4-spec}). At low inclination, the break disappears, while UV photons from the inner flow can still photoionize the BLR, producing strong BELs. 
    \item The Balmer break strength should correlate with the FWHM of the BELs through the inclination effect (Section~\ref{sect:bel}). 
    \item The optical continuum should show weak variability because the large accretion rate and emission-region size imply a long thermal timescale. The BELs should be more variable than the red optical continuum, and the variability of the two components may not be tightly correlated (Section~\ref{sect:var}). 
    \item Because the observed SED is much softer than the intrinsic one, $L_{\mathrm{bol}}/L_{\mathrm{Edd}}$ can substantially underestimate the true Eddington ratio (Section~\ref{sect:ledd}). 
\end{itemize}

According to our model, the LRD phase terminates once the SMBH undergoes significant mass growth and the gas supply can no longer sustain the required high $\dot{m}$. If confirmed, this picture would have important implications for SMBH growth in the early universe (Section~\ref{sect:massgrowth}). In particular, it can accommodate the observed $M_{\mathrm{BH}}$ at $z\sim 7$ even if the accretion rate at the event horizon is only $\dot{m}\sim 10$. Future JWST multi-epoch observations of LRDs can directly test our model predictions, thereby revealing the key mass growth phase in SMBHs.

\begin{acknowledgements}
We thank J.-M. Wang and X. Cao for their helpful discussions, and the anonymous referee for the helpful comments that improved the manuscript. M.Y.S. and S.Y.Z. acknowledge support from the National Natural Science Foundation of China (NSFC-12322303, NSFC-125B2058, NSFC-12533006), the National Key R\&D Program of China (No.~2023YFA1607903), and the Fundamental Research Funds for the Central Universities (20720240152). Y.P.L. was supported by the National Natural Science Foundation of China (grants 12373070, and 12192223). LCH was supported by the National Science Foundation of China (12233001) and the China Manned Space Program (CMS-CSST-2025-A09). W.M.G. acknowledges support from the National Natural Science Foundation of China (NSFC-12221003), the National Key R\&D Program of China (No.~2023YFA1607901) and the China Manned Space Program (CMS-CSST-2025-A13). J.F.W. acknowledges the
National Key R\&D Program of China (grant No. 2023YFA1607904) and NSFC grant 12633004. This manuscript benefited from grammar checking by \texttt{Gemma} and \texttt{Grammarly}. The data products presented herein were retrieved from the Dawn JWST Archive (DJA). DJA is an initiative of the Cosmic Dawn Center (DAWN), which is funded by the Danish National Research Foundation under grant DNRF140. The observations used in the manuscript can be accessed via the following doi: \href{https://doi.org/10.17909/373d-1f39}{10.17909/373d-1f39}.
\end{acknowledgements}

\software{astropy \citep{Astropy2013, Astropy2018, Astropy2022}, Matplotlib \citep{matplotlib}, Numpy \citep{2020NumPy-Array}, Scipy \citep{2020SciPy-NMeth}}

\bibliography{ref.bib}{}
\bibliographystyle{aasjournalv7}

\appendix

\section{Disk vertical structure}\label{appendix:vs}
We solve the vertical structure of the accretion disk by solving the hydrostatic equilibrium, energy balance, and radiative transfer equations. These equations are given by
\begin{equation}
    \label{eq:hydro}
    \frac{dP_{\mathrm{tot}}}{dz} = -\rho \Omega_{\mathrm{K}}^2 z,
\end{equation}
\begin{equation}
    \label{eq:energy}
    \frac{dF}{dz} = q_{\mathrm{rad}}=q_{\mathrm{vis}}-q_{\mathrm{adv}},
\end{equation}
\begin{equation}
    \label{eq:rad}
    \frac{dT}{dz} = -\frac{3\kappa_{\mathrm{R}}\rho F}{16\sigma_{\mathrm{SB}}T^3},
\end{equation}
where $P_{\mathrm{tot}}$ is the total pressure, $\Omega_{\mathrm{K}}$ is the Keplerian angular velocity, $F$ is the radiative flux, $q_{\mathrm{rad}}$ is the radiative cooling rate, $q_{\mathrm{vis}}$ is the viscous heating rate, $q_{\mathrm{adv}}$ is the advective cooling rate, and $z$ is the vertical distance measured from the disk midplane. These equations can be regarded as the generalized version of the thin disk model in \citet{LiuHP2025}. Here, we assume that radiation pressure dominates over gas pressure ($P_{\mathrm{tot}}=P_{\mathrm{rad}}$), which is valid in the outer regions of the SLIM disk. For radiative transport, we neglect convection; this assumption should be tested by future simulations. The boundary conditions are set as $F(z=0)=0$, $F(z=H)=\sigma_{\mathrm{SB}}T_{\mathrm{eff}}^4$, and $P_{\mathrm{rad}}(z=H)=2\sigma_{\mathrm{SB}}T_{\mathrm{eff}}^4/(3c)$, where $H$ is the disk scale height. Following \cite{SSD}, we adopt the $\alpha$-viscosity prescription,
\begin{equation}
    \label{eq:alpha}
    2\alpha \int_{0}^{H} P_{\mathrm{tot}}dz = \frac{1}{2\pi}\Omega_{\mathrm{K}}\dot{M}f,
\end{equation}
where $\alpha$ is the dimensionless viscosity parameter. In our simulations, we adopt a fiducial value of $\alpha=0.1$ unless otherwise specified. These equations can also be applied to an SSD by fixing $q_{\mathrm{adv}}=0$. Note that Eq.~\ref{eq:energy} ensures the vertically integrated radiative cooling rate, $\int_{0}^{H}q_{\mathrm{rad}}dz=F(z=H)-F(z=0)=\sigma_{\mathrm{SB}}T_{\mathrm{eff}}^4$, is smaller than (for the SLIM solution; see Eq.~\ref{eq:T_SLIM}) or identical to (for the SSD solution; see Eq.~\ref{eq:T_SSD}) the vertically integrated viscous heating rate $\int_{0}^{H}q_{\mathrm{vis}}dz=(3GM_{\mathrm{BH}}\dot{M}f)/(8\pi R^3)$. We emphasize that, although our simplified adoption of a constant $\dot{M}$ from the outer to the inner regions appears inconsistent with the scenario proposed in Section~\ref{sect:idea} (where the inner wind decelerates and rejoins the inflow at larger radii), the emergent SED viewed at high inclination is expected to be insensitive to this simplification, because the emission from the inner regions is shielded by the outer regions.
Eqs.~\ref{eq:hydro}--\ref{eq:alpha} are numerically challenging to solve due to the strong coupling among the equations and the extreme sensitivity of the opacity to temperature. We aim to obtain an approximate solution by following the methodology presented in the first appendix of \cite{LiuHP2025}. The key procedure is to first solve the equations by assuming a constant opacity (e.g., $\kappa_{\mathrm{R}}=\kappa_\mathrm{es}=0.34\ \mathrm{cm^2\ g^{-1}}$) to obtain the vertical structures of $\rho(z)$, $T(z)$, and $F(z)$. Then, we fix the profiles of $\rho(z)$, $T(z)$, and $F(z)$ and allow the normalization factor of the $\rho(z)$ profile to be a free parameter. Here, we again assume that $q_{\mathrm{rad}}/q_{\mathrm{vis}}$ is independent of $z$ and that $q_{\mathrm{vis}}\propto \kappa_{\mathrm{R}}\rho \tau^{-0.5}$ \citep{Hirose2009,LiuHP2025}. We can then solve Eq.~\ref{eq:alpha} again with the realistic opacity table to obtain the normalization factors.

\section{Radiative transfer in the disk}\label{appendix:RT}
We have developed a numerical framework to compute the emergent spectral energy distributions (SEDs) of accretion disks by solving the frequency-dependent radiative transfer equation in a two-dimensional $(r,z)$ configuration. The vertical structures of gas density $\rho(r,z)$ and temperature $T(r,z)$ are pre-calculated with the hydrostatic equilibrium, radiative energy transport, thermal equilibrium, and $\alpha$--viscosity prescription as detailed in Appendix~\ref{appendix:vs} and Figure~\ref{fig3-vs}. As the original data are non-uniformly sampled, we use interpolation in logarithmic space to map these quantities onto a uniform global grid. The radial grid spans from $r_\mathrm{min}=10R_\mathrm{S}$ to the self-gravity stability radius $r_\mathrm{max}=R_\mathrm{stable}$, which is defined as Toomre $Q(R_\mathrm{stable})=\Omega_\mathrm{K}^2/2\pi G \rho \approx1$. The vertical grid covers the range $0\le z \le Z_\mathrm{max}$, where $Z_\mathrm{max}$ is defined as the maximum vertical extent of the disk structure across the entire radial range. For regions outside the disk surface, i.e., $Z_\mathrm{surface}\le z \le Z_\mathrm{max}$, we add a artificially small density of $10^{-30}\ \mathrm{g\ cm^{-3}}$ and a temperature of $15\ \mathrm{K}$. The frequency-dependent absorption opacities ($\kappa_\mathrm{abs,\nu}$) and electron scattering opacities ($\kappa_\mathrm{es}$) are obtained from a high-resolution opacity table with a low-metallicity $Z=0.005Z_\odot$ \citep{nupac, Morag2026}. For the low-density regions, the true absorption and electron scattering opacities are fixed to be $10^{-27}\ \mathrm{cm^2\ g^{-1}}$ and $10^{-29}\ \mathrm{cm^2\ g^{-1}}$, respectively, making their contribution to the total optical depths negligible. 

The radiation field is determined by solving the frequency-dependent radiative transfer equation,
\begin{equation}
    \frac{dI_\nu}{ds}=\chi_\nu(S_\nu-I_\nu),
\end{equation}
where $I_\nu$ is the specific intensity, $S_\nu$ is the source function, $ds$ is the path length along a ray, and $\chi_\nu=\rho(\kappa_\mathrm{abs,\nu}+\kappa_\mathrm{es})$ is the total extinction coefficient. We use a short-characteristics (SC) ray-tracing scheme \citep[e.g.,][]{Mihalas1978} on the $(r,z)$ plane. For a given ray direction $\textbf{n}$, the intensity $I_\nu$ at a grid point is calculated by integrating the source function from an upstream intersection point on the cell boundary. To address the potential exponential gradients in the disk atmosphere, we use logarithmic interpolation for $\chi_\nu$, $S_\nu$, and $I_\nu$ at cell interfaces. The optical depth increment $\Delta \tau$ along a segment $\Delta s$ is computing using a logarithmic average, 
\begin{equation}
    \Delta \tau = \Delta s\frac{\chi_\mathrm{loc}-\chi_\mathrm{up}}{\ln (\chi_\mathrm{loc}/\chi_\mathrm{up})},
\end{equation}
where we assume $\chi=\chi_\mathrm{up}e^{s\cdot\ln(\chi_\mathrm{loc}/\chi_\mathrm{up})/\Delta s}$ along $\Delta s$, with $\chi_\mathrm{up}$ and $\chi_\mathrm{loc}$ denoting the total extinction coefficients at the upstream and local points, respectively. The emergent intensity from a cell is then obtained via the formal solution,
\begin{equation}
    I_\mathrm{loc}=I_\mathrm{up} e^{-\Delta \tau}+\int _0 ^{\Delta \tau} S(\tau')e^{-(\Delta \tau-\tau')} d\tau'=I_\mathrm{up} e^{-\Delta \tau}+S_\mathrm{up}w_0+S_\mathrm{loc}w_1,
\end{equation}
where weights $w_0=\frac{1-e^{-\Delta \tau}}{\Delta\tau}-e^{-\Delta \tau}$ and $w_1=1-\frac{1-e^{-\Delta\tau}}{\Delta \tau}$.

To account for isotropic electron scattering, the source function $S_\nu$ is coupled to the mean intensity $J_\nu=\frac{1}{4\pi}\oint I_\nu d\Omega$, 
\begin{equation}
    S_\nu=(1-\epsilon_\nu)J_\nu+\epsilon_\nu B_\nu(T),
\end{equation}
where $B_\nu(T)$ is the Planck function and $\epsilon_\nu=\kappa_\mathrm{abs,\nu}/(\kappa_\mathrm{abs,\nu}+\kappa_\mathrm{es})$. We solve this coupled system using the Lambda-iteration method. In each iteration, we perform a global sweep of rays across the $(r,z)$ grid for a discrete set of angles $\theta$ defined by Gauss-Legendre quadrature. We set boundary conditions at the midplane ($z=0$) and the inner radial boundary  ($r=r_\mathrm{min}$) as $I_\nu=B_\nu$. Convergence is achieved when the maximum relative change in the source function falls below $10^{-3}$.

\begin{figure}
\centering
\includegraphics[width=.6\linewidth]{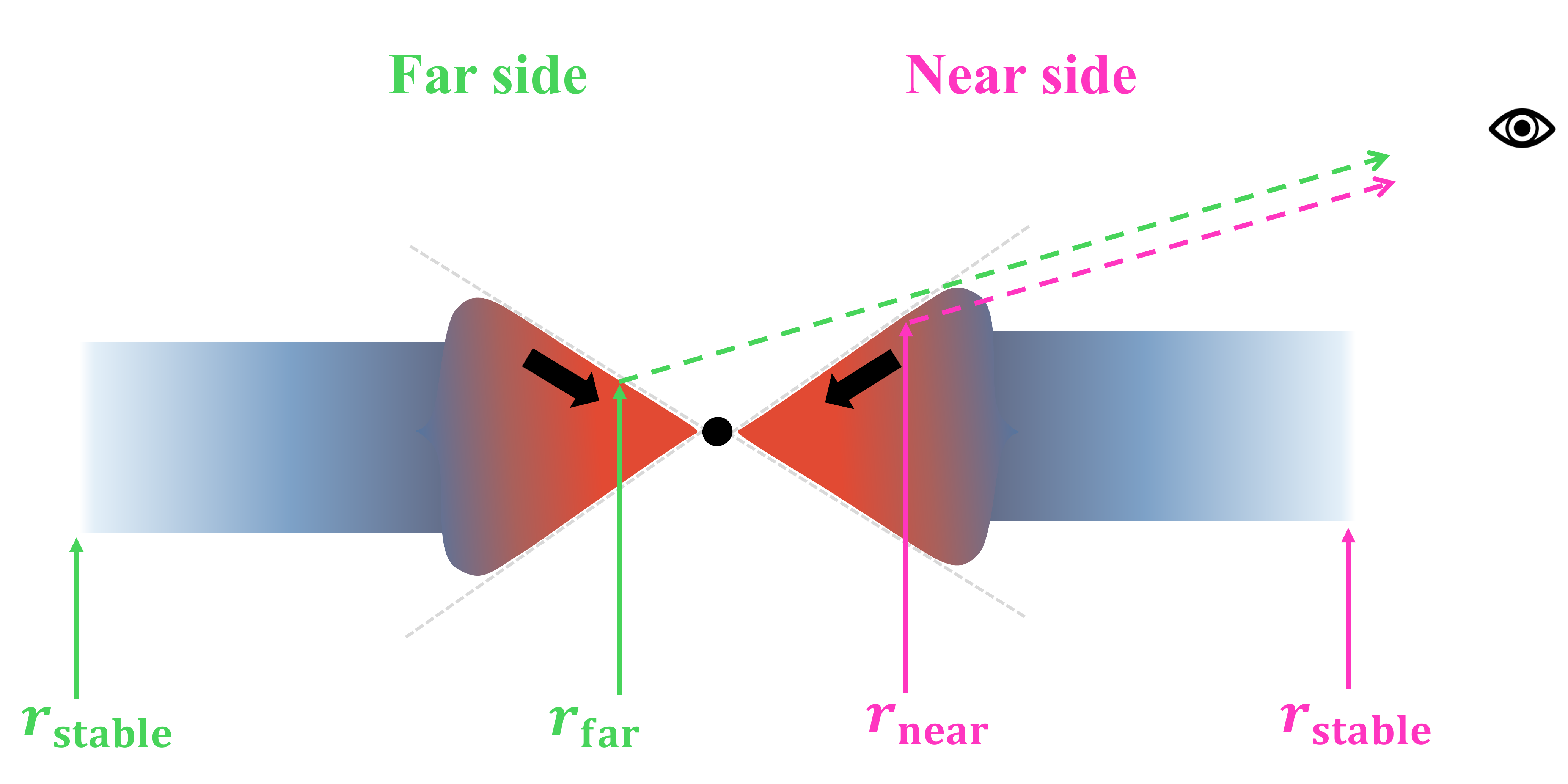}
\caption{Schematic illustration of the disk visible surface. The observable region is bounded by: (i) the near-side innermost radius $r_\mathrm{near}$, where photons escape the advective flow; (ii) the far-side innermost visible radius $r_\mathrm{far}$, determined by the geometric self-shadowing of the inflated SLIM disk; and (iii) the outer boundary $r_\mathrm{stable}$, dictated by the self-gravitational stability limit.
\label{fig-rlimit}}
\end{figure}

Once the source function has converged, we can compute the observable flux for various inclination angles $i_\mathrm{obs}$. For each line of sight, we perform a final ray-trace to extract the emergent intensity $I_\nu(r, i_\mathrm{obs})$ at the $Z_\mathrm{max}$. The observed flux is composed of the near and far sides of the disk (relative to the observer; see Figure~\ref{fig-rlimit}). For the near side of the disk, the innermost radius ($r_\mathrm{near}$) is defined as the radius at which radiation can radially escape the advective flow (see Section~\ref{sect:cont}); this parameter has very limited impact on our final results. For the far side, the visible radius ($r_\mathrm{far}$) is the innermost radius, below which the disk emission is shielded by the near-side SLIM disk. The total monochromatic luminosity $L_\nu$ is then obtained by integrating the intensity over the visible surface area,
\begin{equation}
    L_\nu=\int_{r_\mathrm{near}} ^ {r_\mathrm{max}}\pi I_{\nu,\mathrm{near}}(r, i_\mathrm{obs})r\cos i_\mathrm{obs} dr+\int_{r_\mathrm{far}} ^ {r_\mathrm{max}}\pi I_{\nu, \mathrm{far}}(r, i_\mathrm{obs})r\cos i_\mathrm{obs} dr.
\end{equation}
We expect that $I_{\nu, \mathrm{far}}(r, i_\mathrm{obs})=I_{\nu,\mathrm{near}}(r, -i_\mathrm{obs})$. The isotropic luminosity is defined as $4\pi \nu L_\nu$ to compare with observations.

\begin{figure}
\centering
\includegraphics[width=.8\linewidth]{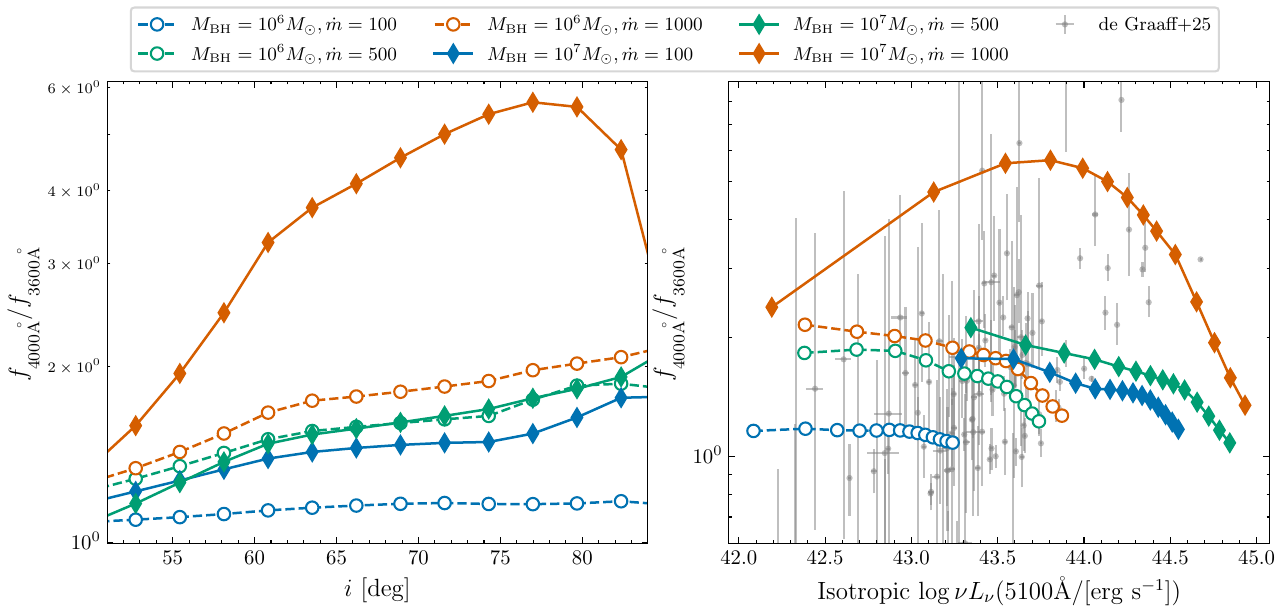}
\caption{Same as Figure~\ref{fig6-break}, but with a viscosity parameter of $\alpha=0.01$. The modification to the vertical structure of the disk, driven by the change in viscosity, results in a more pronounced Balmer break strength. 
\label{fig-break-2}}
\end{figure}

\section{Balmer break tests}\label{appendix:break}
We investigate the sensitivity of the Balmer break strength to the viscosity parameter by comparing our fiducial model ($\alpha=0.1$) with a lower viscosity case ($\alpha=0.01$). As shown in Figure~\ref{fig-break-2}, a smaller viscosity parameter results in a more pronounced Balmer break by changing the disk vertical structure.

\begin{figure}
\centering
\includegraphics[width=.5\linewidth]{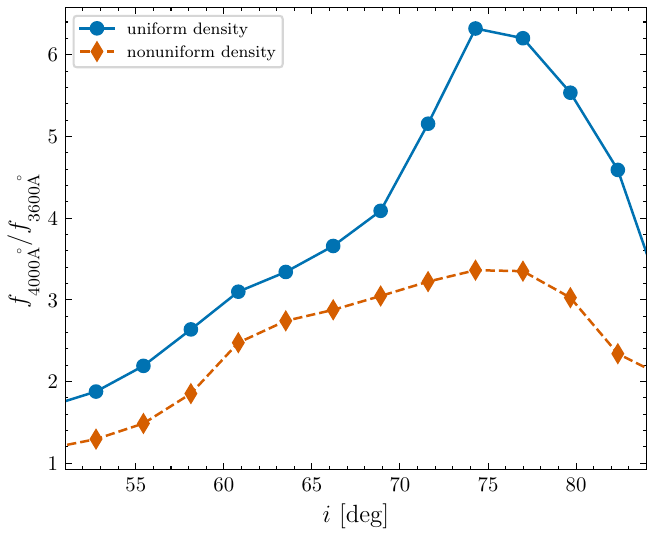}
\caption{Comparison of Balmer break strength between models assuming a simplified uniform vertical gas density versus a realistic nonuniform vertical density profile. Orange diamonds denote the results for nonuniform vertical density (the same markers in Figure~\ref{fig6-break}), while the blue dots represent the simplified, uniform density case. Even under the simplified assumption of uniform density, the Balmer break remains prominent. Consequently, while the vertical structures derived in Section~\ref{sect:balmer} omit certain physical complexities, the fundamental prediction of the Balmer break is robust. 
\label{fig-density}}
\end{figure}

We also compare the nonuniform vertical density profile with a simplified uniform case to test the robustness of the Balmer break. To do so, we enforce a constant density ($\rho = \mathrm{const}$) in Eq.~\ref{eq:hydro}, re-calculate the vertical structure, and subsequently derive the emergent SED. As shown in Figure~\ref{fig-density}, the Balmer break remains prominent under the extreme assumption of a uniform vertical density distribution.

\end{document}